\documentclass[twocolumn,showpacs,prc,superscriptaddress,floatfix,nofootinbib]{revtex4}   
\usepackage{graphicx}

\newcommand{ \be }{\begin{eqnarray}}       
\newcommand{ \ee }{\end{eqnarray}}       
\newcommand{ \la }{\langle}       
\newcommand{ \ra }{\rangle}

\newcommand{ \mean }[1]{\left\langle #1 \right\rangle}   

\def\snn{$\sqrt{s_{NN}}$}
\def\T{$\cal T$}
\def\P{$\cal P$}
\def\CP{$\cal CP$}

\newcommand{\cum}[1]{\left< \left< {#1} \right> \right>}
\newcommand{ \psirp }{\Psi_{RP}}
\newcommand{ \phia }{\phi_{\alpha}}
\newcommand{ \phib }{\phi_{\beta}}
\newcommand{ \corr }{\mean{\cos(\phia+\phib-2\psirp)}}   
\usepackage{color}   
\definecolor{orange}{cmyk}{0.,0.353,1.,0.}    % orange
\definecolor{dgreen}{cmyk}{1.,0.,1.,0.4}	% dark green

\usepackage{fixmath} % bold math headlines          
\usepackage{dcolumn} % Align table columns on decimal point          
\usepackage{bm} % bold math          

\begin{document}

\title{       
%%%%%%%%%%%%%%%%%%%%%%%%% T I T L E %%%%%%%%%%%%%%%%%%%%%%%%%%%%
Observation of charge-dependent azimuthal correlations and possible
local strong parity violation in heavy-ion collisions
} 
 
%\author{The STAR Collaboration}   
%\include{star_author_list}   
\affiliation{Argonne National Laboratory, Argonne, Illinois 60439, USA}
\affiliation{University of Birmingham, Birmingham, United Kingdom}
\affiliation{Brookhaven National Laboratory, Upton, New York 11973, USA}
\affiliation{University of California, Berkeley, California 94720, USA}
\affiliation{University of California, Davis, California 95616, USA}
\affiliation{University of California, Los Angeles, California 90095, USA}
\affiliation{Universidade Estadual de Campinas, Sao Paulo, Brazil}
\affiliation{University of Illinois at Chicago, Chicago, Illinois 60607, USA}
\affiliation{Creighton University, Omaha, Nebraska 68178, USA}
\affiliation{Czech Technical University in Prague, FNSPE, Prague, 115 19, Czech Republic}
\affiliation{Nuclear Physics Institute AS CR, 250 68 \v{R}e\v{z}/Prague, Czech Republic}
\affiliation{Institute of Physics, Bhubaneswar 751005, India}
\affiliation{Indian Institute of Technology, Mumbai, India}
\affiliation{Indiana University, Bloomington, Indiana 47408, USA}
\affiliation{University of Jammu, Jammu 180001, India}
\affiliation{Joint Institute for Nuclear Research, Dubna, 141 980, Russia}
\affiliation{Kent State University, Kent, Ohio 44242, USA}
\affiliation{University of Kentucky, Lexington, Kentucky, 40506-0055, USA}
\affiliation{Institute of Modern Physics, Lanzhou, China}
\affiliation{Lawrence Berkeley National Laboratory, Berkeley, California 94720, USA}
\affiliation{Massachusetts Institute of Technology, Cambridge, MA 02139-4307, USA}
\affiliation{Max-Planck-Institut f\"ur Physik, Munich, Germany}
\affiliation{Michigan State University, East Lansing, Michigan 48824, USA}
\affiliation{Moscow Engineering Physics Institute, Moscow Russia}
\affiliation{City College of New York, New York City, New York 10031, USA}
\affiliation{NIKHEF and Utrecht University, Amsterdam, The Netherlands}
\affiliation{Ohio State University, Columbus, Ohio 43210, USA}
\affiliation{Old Dominion University, Norfolk, VA, 23529, USA}
\affiliation{Panjab University, Chandigarh 160014, India}
\affiliation{Pennsylvania State University, University Park, Pennsylvania 16802, USA}
\affiliation{Institute of High Energy Physics, Protvino, Russia}
\affiliation{Purdue University, West Lafayette, Indiana 47907, USA}
\affiliation{Pusan National University, Pusan, Republic of Korea}
\affiliation{University of Rajasthan, Jaipur 302004, India}
\affiliation{Rice University, Houston, Texas 77251, USA}
\affiliation{Universidade de Sao Paulo, Sao Paulo, Brazil}
\affiliation{University of Science \& Technology of China, Hefei 230026, China}
\affiliation{Shandong University, Jinan, Shandong 250100, China}
\affiliation{Shanghai Institute of Applied Physics, Shanghai 201800, China}
\affiliation{SUBATECH, Nantes, France}
\affiliation{Texas A\&M University, College Station, Texas 77843, USA}
\affiliation{University of Texas, Austin, Texas 78712, USA}
\affiliation{Tsinghua University, Beijing 100084, China}
\affiliation{United States Naval Academy, Annapolis, MD 21402, USA}
\affiliation{Valparaiso University, Valparaiso, Indiana 46383, USA}
\affiliation{Variable Energy Cyclotron Centre, Kolkata 700064, India}
\affiliation{Warsaw University of Technology, Warsaw, Poland}
\affiliation{University of Washington, Seattle, Washington 98195, USA}
\affiliation{Wayne State University, Detroit, Michigan 48201, USA}
\affiliation{Institute of Particle Physics, CCNU (HZNU), Wuhan 430079, China}
\affiliation{Yale University, New Haven, Connecticut 06520, USA}
\affiliation{University of Zagreb, Zagreb, HR-10002, Croatia}

\author{B.~I.~Abelev}\affiliation{University of Illinois at Chicago, Chicago, Illinois 60607, USA}
\author{M.~M.~Aggarwal}\affiliation{Panjab University, Chandigarh 160014, India}
\author{Z.~Ahammed}\affiliation{Variable Energy Cyclotron Centre, Kolkata 700064, India}
\author{A.~V.~Alakhverdyants}\affiliation{Joint Institute for Nuclear Research, Dubna, 141 980, Russia}
\author{B.~D.~Anderson}\affiliation{Kent State University, Kent, Ohio 44242, USA}
\author{D.~Arkhipkin}\affiliation{Brookhaven National Laboratory, Upton, New York 11973, USA}
\author{G.~S.~Averichev}\affiliation{Joint Institute for Nuclear Research, Dubna, 141 980, Russia}
\author{J.~Balewski}\affiliation{Massachusetts Institute of Technology, Cambridge, MA 02139-4307, USA}
\author{O.~Barannikova}\affiliation{University of Illinois at Chicago, Chicago, Illinois 60607, USA}
\author{L.~S.~Barnby}\affiliation{University of Birmingham, Birmingham, United Kingdom}
% ...

\collaboration{STAR Collaboration}\noaffiliation

%%%%%%%%%%%%%%%%%%%%%%%%%% A B S T R A C T %%%%%%%%%%%%%%%%%%%%%%
\begin{abstract}
Parity-odd domains, corresponding to non-trivial topological solutions
of  the QCD  vacuum, might  be  created during  relativistic 
heavy-ion
collisions.  These domains are  predicted to lead to charge separation
of  quarks along  the orbital momentum  of the system
created in non-central collisions.    
To study this effect, we investigate a three particle  mixed  harmonics
azimuthal correlator which is a \P-even observable,  but
  directly  sensitive  to  the  charge  separation  effect.
We  report  measurements of this observable using
the  STAR detector  in  Au+Au and  Cu+Cu
collisions  at  $\sqrt{s_{NN}}$=200  and  62~GeV.  
The  results  are  presented as a function of collision centrality, 
particle separation in rapidity,  and particle  transverse momentum.  
A signal  consistent  with several of the theoretical 
expectations  is   detected in  all  four data sets.    
We  compare  our  results  to  the predictions  of existing  
event  generators, and  discuss in  detail  possible contributions  
from other effects  that are not  related to  parity  violation.  
\end{abstract}

\maketitle

%%%%%%%%%%%%%%%%%%%%%%%%%%%%%%%%%%%%%%%%%%%%%%%%%%%%%%%%%%%%%%
\section{Introduction.}

Quantum Chromodynamics (QCD) is widely accepted as the theory
of the strong interaction. 
The perturbative regime, applying to processes with
large momentum transfer, is theoretically calculable 
and  has been extensively tested~\cite{Amsler:2008zzb}. 
On the other hand, the regime in which quarks and gluons interact 
with modest momenta and with an effective coupling constant 
that is too large for perturbation theory to apply,
cannot be reliably calculated by analytic methods. 
Lattice gauge theory is one first principle approach 
which can be used. 
It predicts the existence of a new state of strongly-interacting
 matter at high energy density.
This state has now been observed in high-energy heavy-ion collisions
at the
Relativistic Heavy Ion Collider (RHIC) at Brookhaven National
Laboratory~\cite{WhitePapers}.   

Many interesting features of this new state of matter 
produced in these collisions have been observed. 
In this paper we focus on a new phenomenon  which we refer to as
local strong  parity (\P) violation. 
It is well known that the strong interaction
conserves parity --- meaning that strong interactions do not 
lead to reactions which produce a finite expectation value for 
any parity odd 
(changing sign under parity transformation) 
observable. 
The best evidence for this comes from experiments 
which set limits on the electric dipole moment 
of the neutron~\cite{Baker:2006ts,Pospelov:1999ha}. 
These experiments show that the parameter $\theta$ 
whose nonzero value would describe parity violation 
in QCD must have magnitude less than $10^{-10}$. % requested by Evan 
This limit effectively makes direct global parity violation 
unobservable in heavy-ion reactions.
Our measurement of a similar \P-odd observable is consistent with 
zero at the level of the experimental precision of
$10^{-4}$ (see section~\ref{sec:results}).

The concept of local parity violation at high temperatures or in 
high-energy heavy-ion collisions was discussed by Lee and Wick 
\cite{Lee:1973iz,Lee:1974ma} and
Morley and Schmidt~\cite{Morley:1983wr} and
elaborated by Kharzeev et al.~\cite{Kharzeev:1998kz}.
In a dense highly excited state, gluon fields can produce
configurations, local in space and time,
which cause \P, time reversal \T, and, via the CPT theorem, 
\CP~violating  effects. 
These field configurations form in different ways 
in different events and averaged over many events they would not 
yield a finite expectation value for a \P-odd observable. 
Each  space-time region, 
occupied by such a configuration, is spontaneously produced
with a random sign of \P-violation, which in the theory is determined by 
 the gluonic field topological 
charge\footnote{
The topological charge distinguishes gluonic field configurations that can
not be continuously transformed one into another.
In general it is not expected to be ``neutralized'' and in 
a given event the net topological charge can take non-zero values. 
For a review of topological effects in QCD, 
see~\cite{Shuryak:1997vd,Diakonov:2002fq}}. 
% end footnote 
Field configurations with non-zero topological charge have
 a finite expectation value for
$\mean{{\vec{E}}_{chromo}\cdot {\vec{B}}_{chromo}}$, 
where ${\vec{E}}_{chromo}$ and 
${\vec{B}}_{chromo}$ are the chromoelectric and chromomagnetic fields,
and the average is taken over the region occupied by the
  configuration.
Since the space time symmetries of chromodynamic fields 
are the same as those of electromagnetic fields,
with electric field being a vector and magnetic field being a pseudovector,  
this region is not invariant under \P~(and \T) transformations. 
Quark interactions with such topological gluonic configurations change
the quark chirality leading to asymmetry in the number of left and
right handed quarks: $N_L -N_R =2 n_f Q$, where $n_f$ is the number of
light quark flavors and $Q$ is the topological charge of the gluonic
configuration.
Thus, the gluonic field configurations
with non-zero topological charge
induce the local \P-violating effects.
Different aspects of an experimental detection of this phenomenon 
were discussed in~\cite{Kharzeev:1998kz,Voloshin:2000xf,Finch:2001hs}.

In  non-central collisions such a domain can manifest 
itself via preferential same charge
particle emission along the system
angular momentum~\cite{Kharzeev:2004ey,Kharzeev:2007tn} (see Fig.~1).
Opposite charge quarks would tend to be emitted in opposite
directions relative to the system angular momentum. 
This asymmetry in the emission of quarks 
would be reflected in, for example, an analogous asymmetry
between positive and negative pion emission directions.
This phenomenon is driven by the large (electro-) magnetic 
field produced in non-central heavy-ion
collisions~\cite{Kharzeev:2004ey,Kharzeev:2007jp,Fukushima:2008xe}.
Peak magnetic field strengths can reach levels of the order of
$10^{15}$~T. 
The combined effect of this magnetic field 
(which tends to align the magnetic moments of the quarks with the field) 
and the difference in the number of quarks with positive 
and negative chiralities (which is
induced by their presence in a ``\P-violating bubble'') 
gives rise to the ``Chiral Magnetic Effect''.

\begin{figure}[ht]
  \includegraphics[width=.44\textwidth]{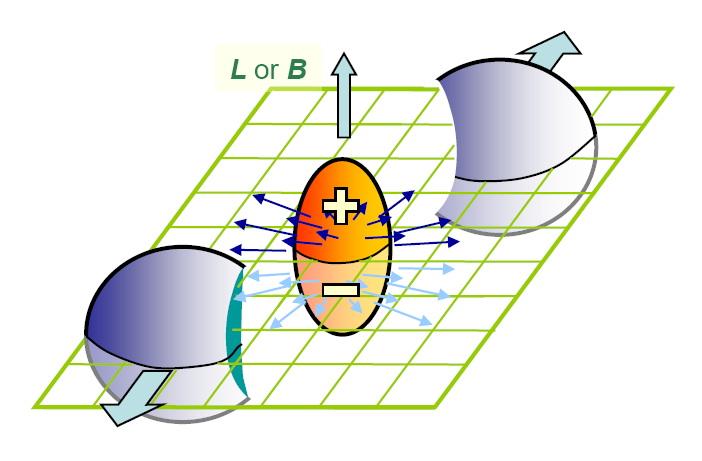}
  \caption{(Color online) 
Schematic view of the charge separation along 
the system orbital momentum. 
The orientation of the charge separation fluctuates in accord
  with the sign of the topological charge.
The direction of the orbital momentum $\bf L$, and that of the magnetic
field $\bf B$, is indicated by an arrow.
}
  \label{fig:charge_sep}
\end{figure}

The same phenomenon can also be described in terms of 
induction of  electric field 
by the (quasi) static magnetic field,
which occurs in the presence of these topologically non-trivial vacuum
solutions ~\cite{Fukushima:2008xe}. 
The induced electric field is parallel to the magnetic
 field, and leads to the charge separation in that direction.
Thus the charge separation  can be viewed as a
non-zero electric dipole moment of the system (see 
Fig.~\ref{fig:charge_sep}).

Depending on the sign of the domain's topological charge, 
positively charged particles will be preferentially emitted 
either along, or in the direction opposite to, the system orbital 
angular momentum, with negative particles
flowing oppositely to the positive particles.
The magnetic field and the angular momentum are normal to
the plane containing the trajectories of the two colliding ions. 
This plane, called the reaction plane, can be found experimentally 
in each collision by observation of the azimuthal distribution 
of produced particles in that event.
                                      
\begin{figure}[ht]
  \includegraphics[width=.44\textwidth]{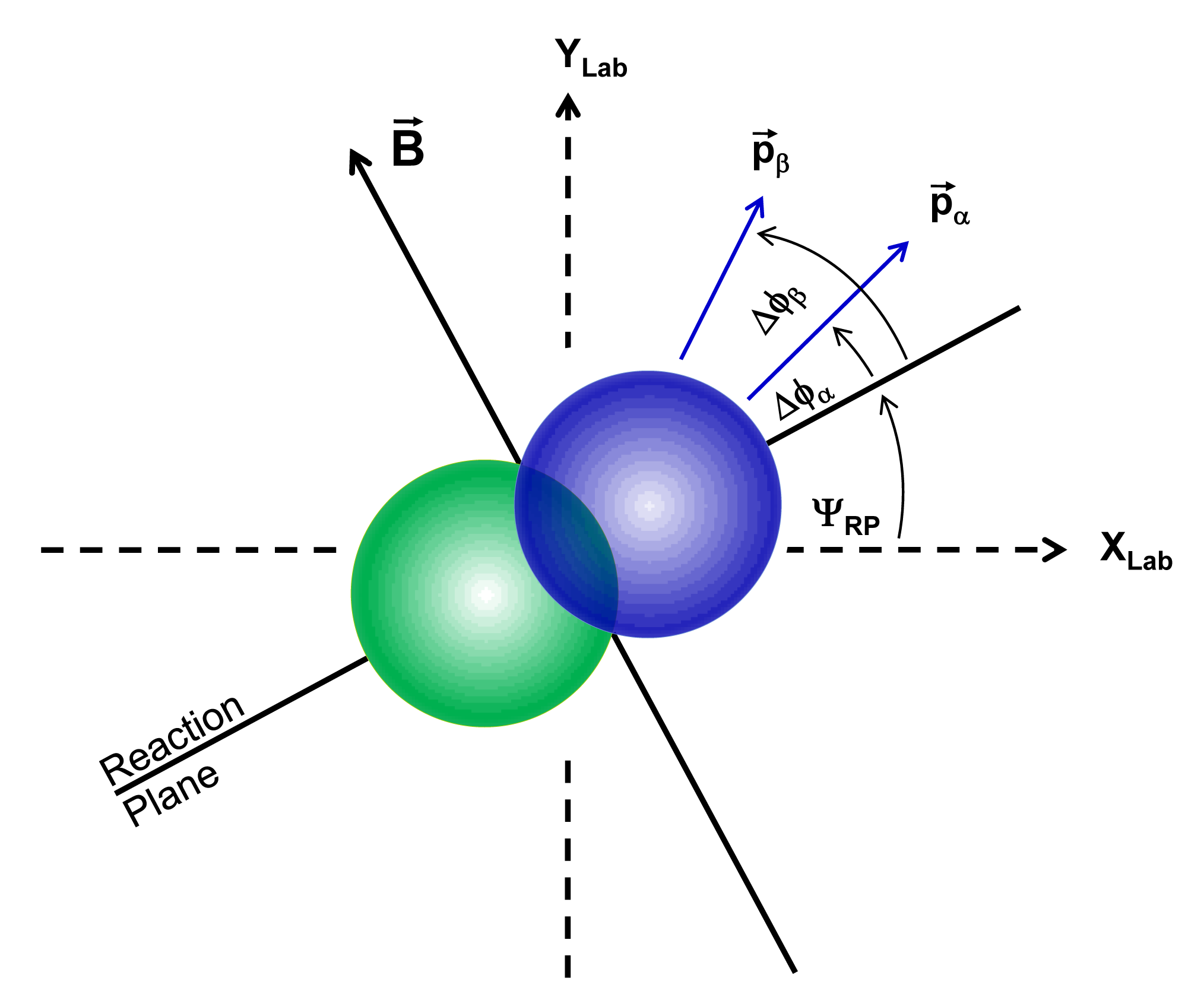}
  \caption{(Color online) 
    Schematic view of the transverse plane indicating the
  orientation of the reaction plane and particle azimuths relative to
  that plane. The colliding nuclei are traveling into and out of the figure.
}
  \label{fig:charge_sep2}
\end{figure}

When two heavy ions collide with a finite impact parameter, the
probability for particles to be emitted in a given azimuthal direction
is often described with a Fourier 
decomposition~\cite{Voloshin:1994mz}:
\be
 \frac{dN_\alpha}{d\phi} &\propto&  1+ 2 v_{1,\alpha} \cos(\Delta \phi)+
2\, v_{2,\alpha} \cos(2\Delta\phi)+... 
\, ,
\label{eq:expansionFlow}
\ee
where $\Delta \phi =(\phi-\psirp)$ is the particle azimuthal direction 
relative to the reaction plane, as shown in Fig. 2.  $v_1$ and $v_2$ are
coefficients accounting for the so-called directed and elliptic flow,
respectively, and $\alpha$ indicates the particle type.  
They depend on the impact parameter of the
colliding nuclei, the particle type ($\pi$, $K$, $p$, ...),
transverse momentum ($p_t$), and pseudorapidity ($\eta$) of the produced
particles.  
For collisions of identical nuclei, symmetry requires $v_1$ to
be an odd function of rapidity 
and $v_2$ to be an even function of rapidity.
Measurements (for a review and references, see~\cite{Voloshin:2008dg})
have found that, at RHIC, $v_1$ is quite small at
mid-rapidity; typically, $|v_1| < 0.005$ for $-1<\eta<+1$.  
In contrast, $v_2$ is found to be sizable and positive.  
In Au+Au collisions at \snn = 200 GeV, for unidentified charged hadrons, $v_2$
reaches 0.25 for $p_t \sim 3$~GeV/c, and 0.06 when integrated over all
$p_t$.

Phenomenologically, the charge separation due to a domain with 
a given sign of the topological charge can be  described 
by adding \P-odd sine terms to the Fourier decomposition 
Eq.~\ref{eq:expansionFlow}~\cite{Voloshin:2004vk}: 
\be
 \frac{dN_\alpha}{d\phi} &\propto&  1+ 2 v_{1,\alpha} \cos(\Delta \phi)+
2\, v_{2,\alpha} \cos(2\Delta\phi)+... 
\nonumber \\
&+&2  a_{1,\alpha} \sin(\Delta \phi) +
2\, a_{2,\alpha} \sin(2 \Delta \phi) +... \, ,
\label{eq:expansion}
\ee
where the $a$ parameters describe the \P-violating effect.
Equation~\ref{eq:expansion} describes the azimuthal distribution of
particles of a given transverse momentum and rapidity and,
like the flow coefficients, $a$ coefficients depend on
transverse momentum and rapidity of the particles. 
In addition, they depend also on the rapidity (position) of the domain.
One expects that only particles close in rapidity 
to the domain position are affected. 
According to the theory, the signs of $a$ coefficients vary 
following the fluctuations in the domain's topological charge.
If the particle distributions are averaged over many events, then
these coefficients will vanish because the distributions are averaged over
several domains with different signs of the topological charge.  
However, the effect of these domains on charged particle correlations will not
vanish in this average, as discussed below.
In this analysis we consider only  the first harmonic
coefficient $a_1$, which is expected to account for most of the
effect although higher harmonics determine the exact shape of the
distribution.  For brevity we will omit the harmonic number,
and write $a_{\alpha}=a_{1,\alpha}$.
The index $\alpha$ takes only two values, $+$ and $-$, for 
positively and negatively charged
 particles respectively. 

The effects of local parity violation
cannot be significantly observed in a single event because of the
statistical fluctuations in the large number of particles,
which are not affected by the \P-violating fields.
The average of $a_\alpha$ over many events, $\mean{a_\alpha}$, 
must be zero. 
The observation of the effect is possible only via correlations,
e.g. measuring $\mean{a_\alpha a_\beta}$
with the average taken over all events in a given event sample.
The correlator $\mean{a_\alpha a_\beta}$ is, however, a \P-even
quantity, and an experimental measurement of this quantity 
may contain contributions from effects unrelated to parity violation.  
The correlator $\mean{a_\alpha a_\beta}$ can be in principle evaluated
via measuring $\mean{\sin \Delta \phia\,\sin\Delta \phib}$
with the average in the last expression taken over all pairs of
particles of a given type from a kinematic region under
study and then over all events.
The problem is that this form of correlator contains also a large 
contribution from correlations not related to the reaction plane 
orientation (such correlations are not accounted for by
Eq.~\ref{eq:expansion}, which is a single particle distribution).
For example, a pair of particles originating from a single jet will 
typically be emitted with a small azimuthal separation.  
These particle pairs will make a positive
contribution to $\mean{\sin\Delta \phia\, \sin\Delta \phib}$, 
even if the parent jets are emitted isotropically relative to 
the reaction plane.  
Therefore, we separate
\be
\mean{\sin\Delta \phia\, \sin\Delta \phib} = \mean{a_\alpha a_\beta}
+B_{out}, 
\label{eq:bout}
\ee
where $\mean{a_\alpha a_\beta}$ is caused by parity violation  and  $B_{out}$ 
(defined by this expression) includes all other 
correlations projected onto the direction perpendicular to
the reaction plane (``out of plane'').  
The effects contributing to $B_{out}$ may
be large and are difficult to estimate reliably.
For this reason, a different correlator was 
proposed~\cite{Voloshin:2004vk}:
\be
\hspace*{-2cm}
& \mean{ \cos(\phia +\phib -2\psirp) } = 
\label{eq:obs1}
\\
&
\mean{\cos\Delta \phia\, \cos\Delta \phib} 
-\mean{\sin\Delta \phia\,\sin\Delta \phib}=
\label{eq:cossin}
\nonumber 
\\ 
& 
[\mean{v_{1,\alpha}v_{1,\beta}} + B_{in}] - [\mean{a_\alpha a_\beta}
+ B_{out}],
\label{eq:v-a}
\ee
where, similarly to Eq.~\ref{eq:bout}, $B_{in}$ is defined via:
\be
\mean{\cos\Delta \phia\, \cos\Delta \phib} = \mean{v_{1,\alpha} v_{1,\beta}}
+B_{in}. 
\label{eq:bin}
\ee
The correlator Eq.~\ref{eq:obs1} represents the
difference between correlations 
of the projections of the particle transverse momentum
unit vectors onto an axis in the reaction plane 
and the correlations of the projections onto an axis
that is out-of-plane or perpendicular to the reaction plane.
The key advantage of using Eq.~\ref{eq:v-a} is that it
removes all the correlations among
particles $\alpha$ and $\beta$ that are not related to the reaction plane 
orientation~\cite{Borghini:2002vp,Adams:2003zg}.

The contribution given by the term $\mean{v_{1,\alpha}v_{1,\beta}}$
can be neglected because directed flow averages to zero in
a rapidity region symmetric with respect to mid-rapidity, as used in this
analysis
and the contribution due to directed flow fluctuations is very small
(see Section~\ref{sec:simulations} for a quantitative estimate).
Equation~\ref{eq:v-a} then implies that by using 
$\corr$ instead of $\mean{\sin\Delta \phia\,\sin\Delta \phib}$,
the background to our measurement of $\mean{a_\alpha a_\beta}$
is now not $B_{out}$, but $[B_{out}-B_{in}]$, where 
$B_{in}$ is the contribution of the in-plane correlations 
which are analogous to $B_{out}$. 
Only the parts of such correlations that depend on azimuthal orientation 
with respect to the reaction plane remain as backgrounds.  
Studies of the various physics contributions to 
$[B_{out}-B_{in}]$ are discussed in detail
in Section~\ref{sec:simulations}.

Based on the current theoretical understanding of the Chiral 
Magnetic Effect one might expect the following features 
of the correlator $\mean{a_\alpha a_\beta}$:
\begin{itemize}
\item 
{\em Magnitude}: The first estimates~\cite{Kharzeev:2004ey}
 predicted a signal of the order
  of $|a|\sim Q/N_{\pi^+}$, where $Q=0,\pm 1,\pm 2, ...$ 
is the net topological charge and $N_{\pi^+}$ is the positive 
pion multiplicity in one unit of rapidity 
-- the expected rapidity scale for correlations due to topological
domains, see below. 
More accurate estimates~\cite{Kharzeev:2007jp} 
including the strength of the magnetic field and topological 
domains production rates, were found to be close to the same number.
It corresponds to values of $|a|$ of the order of $10^{-2}$ for 
mid-central collisions, and
to $10^{-4}$ for the correlator $\mean{a_\alpha a_\beta}$.
\item
{\em Charge combinations}: 
If the particles, after leaving the domain experience
no medium effects (re-interaction with other particles in the system), 
one would expect $a_+=-a_-$.
Thus, in the absence of medium effects, one expects 
$\mean{a_+a_+}=\mean{a_-a_-}=-\mean{a_+a_-} >0$. 
If the process occurs in a dense
medium one needs to account for correlation modifications due to
particle interaction with the medium~\cite{Kharzeev:2007jp}. 
The effect of these modifications is similar to the
modification of the jet-like two-particle correlations 
which experience  strong
suppression of the back-to-back correlations:
$\mean{a_+a_+}=\mean{a_-a_-} \gg -\mean{a_+a_-}$. 
The effect of strong radial flow can further modify this relation such
that the opposite charge correlations can even become positive.
\item
{\em Centrality dependence:} 
Under the assumption that the average size of the 
\P-violating domain does not change with centrality, the correlator
should follow a $1/N$ dependence (typical for any
kind of correlations due to clusters; $N$ is the multiplicity) 
multiplied by a factor
accounting for the variation of the magnetic field.     
The latter is  difficult to predict reliably at present,
other than that it should be zero in perfectly central collisions.
Thus at large centralities the effect should decrease with centrality 
somewhat faster than $1/N$.
\item
{\em Rapidity dependence:} 
The correlated particles are produced in a domain of the order 
of 1~fm, and it is expected that the correlations should have 
a width in $\Delta\eta=|\eta_\alpha-\eta_\beta |$ of the order unity,
as is typical for hadronic production from clusters~\cite{Foa:1975eu}.
\item
{\em Transverse momentum dependence:}
Local parity violation is non-perturbative in nature and 
the main contribution to the signal should ``come
from particles which have transverse momentum smaller than
1~GeV/c''~\cite{Kharzeev:2007jp}.
The actual limits might be affected by the radial flow.
\item
{\em Beam species dependence:} 
The effect should be proportional to the square $Z^2$ of the
nuclear charge,  but the
atomic number $A$ dependence is not well understood.
One qualitative prediction is that the suppression of the 
back-to-back correlations should be smaller in collisions of
lighter nuclei. 
\item
{\em Collision energy dependence:}
The effect might be stronger at lower energies, as the time integral
of the magnetic field is larger.
At the same time,
the charge separation effect is expected to depend strongly on
deconfinement
and chiral symmetry restoration~\cite{Kharzeev:2007jp}, 
and the signal might be greatly suppressed or completely absent
at an energy below that at which a quark-gluon plasma can be formed.
\end{itemize}

The main systematic uncertainty in application of the 
correlator Eq.~\ref{eq:obs1}
to measurements of anisotropies in particle production with
respect to the reaction plane, is due to processes when 
particles $\alpha$ and $\beta$ are products of a cluster 
(e.g. resonance, jet) decay, and the cluster itself exhibits
elliptic flow~\cite{Borghini:2001vi,Adams:2003zg}.
Detailed discussion of this and other effects which could mimic the
effect of local strong parity violation in 
experimental measurements is presented in 
section~\ref{sec:simulations}.

In this paper, we report our measurements of  
correlators shown in Eqs.~\ref{eq:obs1}-\ref{eq:v-a}
and present systematic studies of the background effects 
that affect the measurements. 
Section~\ref{sec:setup} discusses the
experimental setup, while section~\ref{sec:method} discusses 
the observables and the
methods for estimating the reaction plane angle
and corrections for finite reaction plane resolution.  
Sections~\ref{sec:detector_effects} and \ref{sec:syst} present 
the data and a
discussion of systematic
 effects that can affect the measurements.  
Our main results, and how they systematically change with system size, 
centrality, particle transverse momentum, and separation in rapidity,
are presented in Section~\ref{sec:results}. 
Physics backgrounds that can
mimic the \P-violating effect are discussed in 
section~\ref{sec:simulations}.

%=================================================================
\section{Experimental Setup and Data Taking}
\label{sec:setup}

The data were collected with the STAR detector at Brookhaven National
Laboratory during the 2004 and 2005 runs. Collisions of Au+Au and Cu+Cu
beams were recorded at $\sqrt{s_{NN}}=200$ and 62 GeV incident energies; 
for a total of four beam-energy
combinations. Charged particle tracks were reconstructed in a cylindrical
Time Projection Chamber (TPC)~\cite{Ackermann:2002ad,Anderson:2003ur}. 
The TPC is a 4.2~m long barrel
with a 2~m radius which was operated in a solenoidal magnetic 
field of 0.5~T.
The TPC detects charged tracks with pseudorapidity $|\eta| < 1.2$ 
and $p_t > 100$~MeV/c with 
an absolute efficiency that ranges from 80 to 90\%.
The TPC is nearly azimuthally
symmetric and records tracks at all azimuthal angles;
however, sector boundaries and other regular detector features 
are responsible for an approximately 10\% loss of particles 
due to the finite acceptance of the detector.  
Track merging, and other tracking artifacts that depend on track
density, can cause an additional 0-10\% loss of reconstructed tracks; so the
overall efficiency is typically 85\% per event.

The TPC's pseudorapidity coverage of an event is supplemented 
by two cylindrical
and azimuthally symmetric Forward Time Projection Chambers (FTPC). 
The \mbox{FTPCs} are placed in the forward and backward direction 
relative to the main TPC and cover pseudorapidity intervals 
 $2.7 < | \eta | < 3.9$~\cite{Ackermann:2002yx}.  
In the most forward direction, STAR has two Zero Degree 
Calorimeter - Shower Maximum Detectors 
(ZDC-SMD)~\cite{Adams:2005ca,Adler:2001fq} which are sensitive 
to the directed flow of neutrons in the beam rapidity regions.

A minimum bias trigger was used during data-taking. 
Events with a primary vertex within 30~cm along the beam line from 
the center of the main TPC were selected for the analysis. 
Standard STAR software cuts were
applied to suppress pile-up and  other malformed events or tracks. 
The results presented here are based on 14.7M Au+Au
and 13.9M Cu+Cu events at the center of mass energy of a nucleon pair
\snn=200~GeV, and 2.4M Au+Au and 6.3M Cu+Cu events at \snn=62~GeV. 
The data were taken with the magnetic field in the
Full Field (FF), and Reverse Full Field (RFF) configurations, with the
strength of the magnetic field at 0.5~T. 
The centrality of the collision is
determined according to the reference multiplicity (refMult), which is the
recorded multiplicity of charged particles in $|\eta| < 0.5$ that satisfy
specific track quality cuts.

The correlations are reported in the pseudorapidity region $|\eta| < 1.0$
covered by the main TPC.  
For this analysis, the tracks in the TPC are required to have 
$p_t > 0.15$~GeV/c. 
For the results integrated over transverse momentum we also
impose an upper cut of $p_t < 2$~GeV/c. 
Standard STAR track quality cuts are applied: a minimum of 15 tracking 
points are required for a track to be considered good.
The ratio of the number of hit points on a track to 
the maximum possible given the track geometry is required to
be greater than 0.52 to avoid the effects of track splitting. 
The data with reverse magnetic field  were used 
to assess systematic effects, as the biases for positive and
negative charged particles interchange.
The final results reported here are averaged over both field polarities.

We use particle identification via specific energy loss ($dE/dx$) in the
volume of the TPC to reject electrons as a check that the signal 
we present is determined by hadron production.

%============================================================
\section{Method}
\label{sec:method}

In practice the reaction plane angle for a given collision
is not known.
In order to evaluate the correlator defined in Eq.~\ref{eq:obs1},
one estimates the reaction plane with the so-called
event  plane reconstructed from particle azimuthal 
distributions~~\cite{Poskanzer:1998yz}. 
For the event plane determination one can use particles 
found in the same detector that is used to detect
particles $\alpha$ and $\beta$ (in our case STAR's main TPC)  
or different detectors (we have used the STAR FTPCs and the ZDC-SMD). 
The second order event plane (determined by the 
second harmonic modulation in particle distribution) is sufficient 
for this study.
We make use of the large elliptic flow measured at 
RHIC~\cite{Ackermann:2000tr} to determine the event plane 
from particle distributions in the main and Forward TPCs.
When using the ZDC-SMD for event plane reconstruction, the first-order 
event plane can be determined through the measured directed flow of
spectator neutrons.

In the three-particle correlation technique, the explicit determination of the
event plane is not required; instead, the role of the event plane
is played by the third particle that enters the  correlator 
with double the azimuth~\cite{Poskanzer:1998yz,Borghini:2001vi,Adams:2003zg}.
Under the assumption that particle $c$ is correlated with particles
$\alpha$ and $\beta$ only via common correlation to the reaction
plane, we have: 
\be
\la \cos(\phi_a +\phi_\beta -2\phi_c) \ra 
=
\la \cos(\phi_a +\phi_\beta -2\psirp) \ra \, v_{2,c}, 
\label{e3p}
\ee  
where $v_{2,c}$ is the elliptic flow value of the particle $c$.
We check this assumption by using particles $c$ from different
detectors and exhibiting different elliptic flow.
We also study the effect of using only positive or only
negative particles to determine the event plane and compare
the results obtained with different field polarities in our
estimates of the systematic uncertainties.
All the correlators presented in this paper have been calculated 
by first averaging over all  particles under study in a given 
event and subsequently averaging  the results over all events 
in a given  event sample.

The STAR TPCs have quite uniform azimuthal acceptance. 
Nevertheless, TPC sector boundaries, malfunctioning electronics, etc., 
may introduce biases in the analysis, in particular as 
the acceptance for positive and negative particles is different.
To avoid these effects we use a recentering 
procedure~\cite{Poskanzer:1998yz} in which we substitute:
$\cos\phi \rightarrow \cos\phi - \mean{\cos\phi}$ and
$\sin\phi \rightarrow \sin\phi - \mean{\sin\phi}$
and similarly for the second harmonic.
The typical values of $\mean{\cos\phi}$ and $\mean{\sin\phi}$
for the tracks in the main TPC 
are $\lesssim 0.003$, 
but for high $p_t$ particles
and the most central collisions could go as high as 1.5\%.  
In the FTPC region, the typical correction is of the order of a few
percent.
The validity of the recentering method can be verified by calculating
three-particle cumulants~\cite{Borghini:2002vp,Selyuzhenkov:2007zi}:
\be
& 
\mean{ \cos(\phia +\phib -2\phi_c) } =
\nonumber
\\
& 
=
\Re \{\cum{u_\alpha u_\beta v_c^2} 
+ \mean{u_\alpha u_\beta} \mean{v_c^2}
+ \mean{u_\alpha v_c^2} \mean{u_\beta}
\nonumber
\\
&
+ \mean{u_\beta v_c^2} \mean{u_\alpha}
-2\mean{u_\alpha}\mean{u_\beta}\mean{v_c^2}\},
\label{eq:cumulant}
\ee 
where we use notations $u=e^{i\phi}$ and $v=u^*=e^{-i\phi}$.
$\Re \{...\}$ denotes the real part,
and double angle brackets denote cumulants.
In the case of perfect
acceptance, the cumulant $\cum{ \cos(\phia +\phib -2\phi_c) }$
coincides with the correlator
$\la \cos(\phi_\alpha +\phi_\beta -2\phi_c) \ra $.
As can be seen from  Eq.~\ref{eq:cumulant}, to account for 
the acceptance effect it is sufficient to perform a recentering 
procedure. 
All results presented in this paper have been corrected for
acceptance effects, where applicable, by this method.
The cumulant Eq.~\ref{eq:cumulant} can be calculated directly 
by correcting the three-particle correlator
with the corresponding products of two and single particle averages.
We have compared the results obtained by directly calculating cumulants 
with the results obtained by the recentering method and found
them to be consistent.
Because the detector acceptance varied during the 
 period of data
taking, we perform the corresponding correction run-by-run, separately
for positive and negative particles, 
and for each centrality bin. 
We also account for the acceptance dependence on particle
pseudorapidity and transverse momentum. 
We consider separately 
the East ($\eta<0$) and West ($\eta>0$) FTPCs. 
We have found that the corrections do not depend significantly on the 
collision vertex position along the beam line.

%===============================================================
\section{Detector effect studies}
\label{sec:detector_effects}

\begin{figure}[ht]
\centerline{
\includegraphics[width=0.5\textwidth]{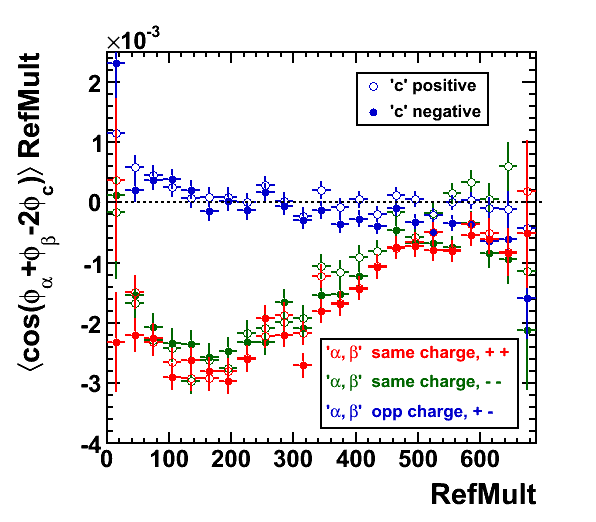}}
\centerline{(a)}
\centerline{
  \includegraphics[width=0.5\textwidth]{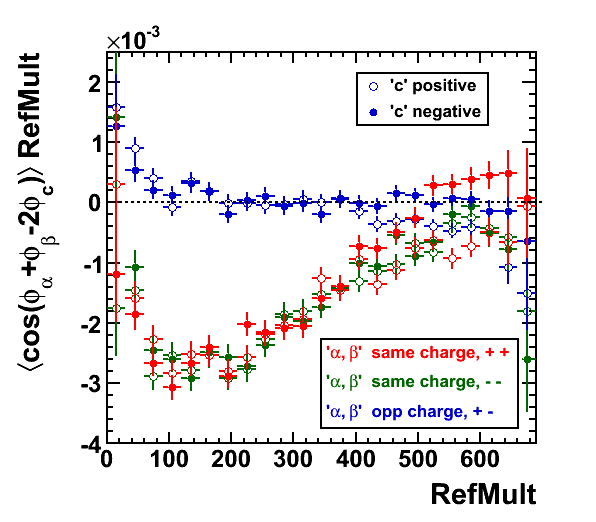}}
\centerline{(b)}
  \caption{(Color)
  $\mean{\cos(\phia +\phib -2\phi_c)}$ as a function of reference
  multiplicity for different charge combinations, before
corrections for acceptance effects. In the legend 
  the signs indicate the charge of particles $\alpha$,
  $\beta$, and $c$. 
   The results shown are for Au+Au collisions at 200 GeV obtained in (a)
  the Reversed Full Field, and (b)  the Full Field configurations.   
}
  \label{fig:no_recenter}
\end{figure}

\begin{figure}[ht]
\centerline{
\includegraphics[width=0.5\textwidth]{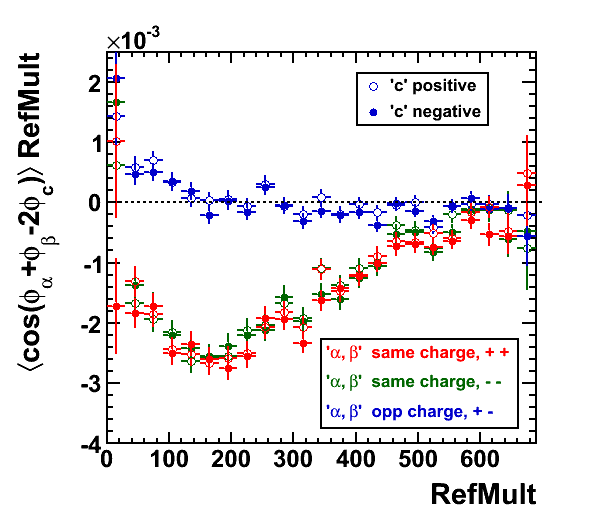}
}
\centerline{(a)}
\centerline{
  \includegraphics[width=0.5\textwidth]{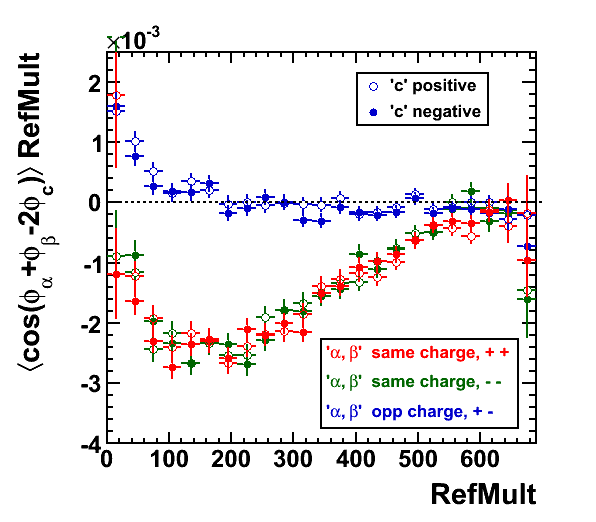}
}
\centerline{(b)}
  \caption{(Color) 
Same as Fig.~\ref{fig:no_recenter} after correction for
  acceptance effects.
}
  \label{fig:recenter}
\end{figure}

Figure~\ref{fig:no_recenter} shows the three-particle correlator, 
Eq.~\ref{e3p}, as a function
of reference multiplicity in Au+Au collisions at \snn=200~GeV for two
field polarities before the recentering procedure.
All three particles are from main TPC region, $|\eta|<1.0$.
Figure~\ref{fig:recenter} shows 
results for the same correlator after correction.
The correlator
has been scaled by the reference multiplicity for clarity at
high centralities, where the absolute values of the signal are small.
These figures are intended only to illustrate the effect of the
recentering; for that reason and also to have finer binning in
centrality, we plot the correlator
 directly versus reference multiplicity.
All other results are presented as a function of the fraction of the
total interaction cross section (which is calculated taking into
account the track, event vertex reconstruction, and 
trigger inefficiencies).   
The acceptance effects are most noticeable for central collisions, 
where the signal is small; there is a slight difference in results 
depending on whether the third ($c$) particle
is positive or negative and the difference changes sign 
depending on the polarity of the magnetic field.
This difference disappears after the acceptance correction.
Results for particles $\alpha$ and $\beta$ being both positive or
both negative are consistent within statistical errors, and later we
combine them  as same charge correlations.
As expected for the case when particles   $\alpha$ and
$\beta$ are correlated to the particle $c$ only via common correlation to the
reaction plane, 
the results do not depend on the charge of the particle $c$.


The acceptance effects are larger in the average correlation, 
$\mean{\cos(\phia-\phib)}$, than 
in the correlator $\mean{\cos(\phia+\phib-2\psirp)}$, 
because the latter represents
the difference in correlations projected onto the reaction plane and
to the direction normal to the reaction plane.  
As the reaction plane is uniformly distributed in azimuth, 
many of the possible acceptance effects average out to zero. 

\begin{figure}[ht]
  \includegraphics[width=.47\textwidth]{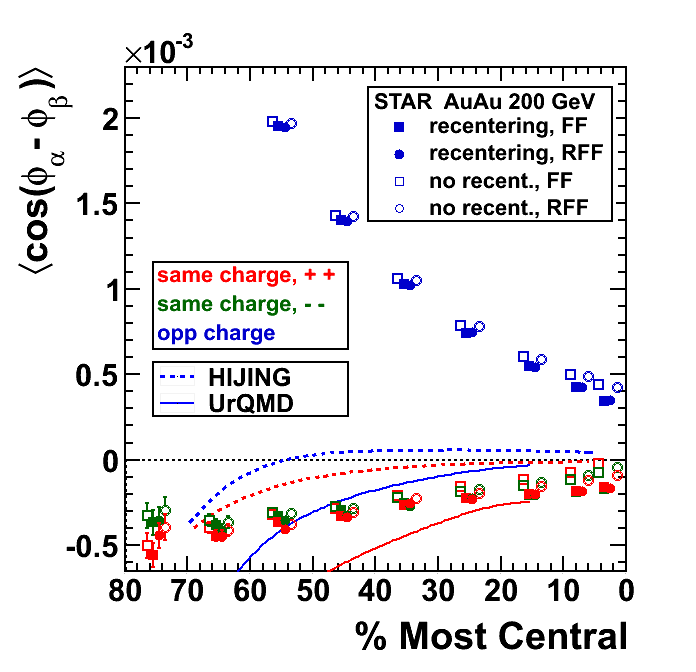}
  \caption{ (Color)
 $\mean{\cos(\phia -\phib)}$ as a function of centrality
   for different charge combinations and FF and RFF  configurations.   
   The data points corresponding to different charge and field
  configurations are slightly shifted in the horizontal direction
 with respect to each other for clarity.
 The error bars are statistical. 
Also shown are model predictions described in Section~\ref{sec:simulations}.
}
  \label{fig:uvAu200}
\end{figure}

Figure~\ref{fig:uvAu200} presents the correlator
$\mean{\cos(\phia-\phib)}$ for different charge combinations from the
Au+Au 200~GeV data obtained with FF and RFF magnetic field settings as a
function of collision centrality. 
In this figure and later in the paper, the centrality is quantified by the 
fraction of the total interaction cross section, with the 
centrality bins corresponding to (ordered from most to least central)
0-5\%, 5-10\%, 10-20\%, ..., 70-80\%
 of the most central collisions.
The points are plotted at the middle of the bin, not reflecting possible 
small biases due to higher weight of events with larger multiplicity within 
the bin. 
Before acceptance corrections are applied, $(+,+)$ correlations
are slightly different from $(-,-)$ correlations, 
with the difference changing sign in different field orientations. 
After the correction, the results from different field polarities
coincide with each other.

We have performed several additional checks to ensure that the signal is
not due to detector effects. 
High accelerator luminosity
leads to significant charge buildup in the TPC, which leads to
distortions in the recorded track positions, affecting the
reconstructed momenta.
We have compared the results obtained from the 2002 RHIC run 
(a low luminosity run), with results from 2004-2005 
divided into high and low luminosity events (selection is based on ZDC
coincidence rate). 
All three data samples yield the same signal within
statistical uncertainties. 

The acceptance of the detector depends weakly on the position of
the event vertex relative to the center of the TPC. 
We   applied  the  acceptance   corrections  differentially
according  to the event  vertex position,  and explicitly
checked  the dependence  of  the  signal on  the  vertex position.  
No dependence has been found.

The  main TPC  consists  of  two parts  which  are separated  by
a central  membrane.
A particle track will occasionally cross the central membrane,
and be separately reconstructed  in each half-barrel  of the TPC.   
These two track parts can be displaced one with respect to 
the other. 
In order to check that this effect  does not contribute to the signal, we
calculated the correlator using only tracks that do not cross the membrane.  
Taking into  account  the  signal  dependence  on  the  track  
separation  in pseudorapidity, the observed signal was found to be 
consistent with  the signal obtained  without such  a requirement.
 
Tracks in the  TPC are characterized by the  distance of closest
approach ($dca$),  the distance between  the projection of  the track,
and  the event  vertex.  Particles  originating from  weak
decays ($\Lambda$, $K_s$, etc.) can have larger $dca$s than 
the direct primary particles we are studying.  
We compared the results obtained with a cut $dca<1$~cm to
those  of  $dca<3$~cm,  and  found only  negligible  differences  with
a somewhat larger signal  (of the order of the statistical
error) for tracks with $dca<1$~cm.

The correlator  used in this  analysis is the  difference between
the  correlations projected  onto the  reaction plane  and the
correlations projected  onto the  direction normal to  the reaction
plane.  
The correlator  calculated by  projecting  onto an
axis rotated by $\pi/4$ relative to the reaction plane 
should only be non-zero due to detector effects.  
We have explicitly calculated the correlator in this rotated  
frame and found it to be zero within statistical error.

\begin{figure}[ht]
  \includegraphics[width=0.5\textwidth]{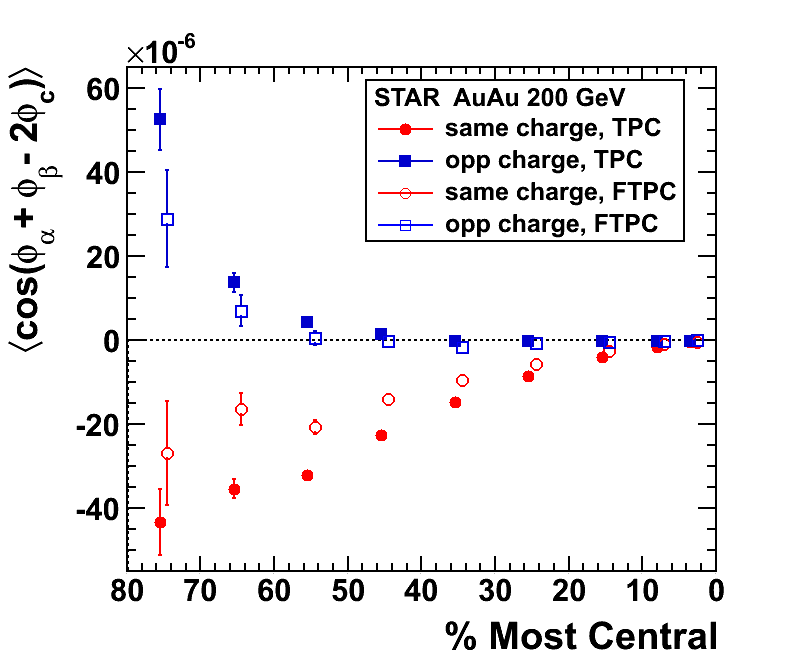}
	\centerline{(a)}
  \includegraphics[width=0.5\textwidth]{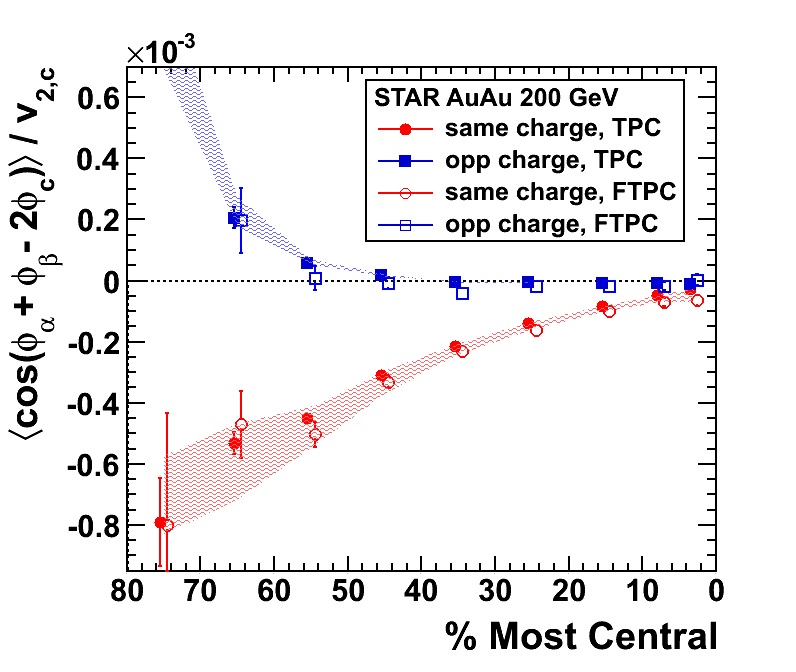}
	\centerline{(b)}
  \caption{ (Color online)
    (a) A comparison of the correlations obtained by
    selecting the  third particle  from the main  TPC or
    from  the Forward  TPCs.   
    (b)  
    The  results after scaling by the flow of the  third  particle.
    The shaded areas represent the uncertainty from $v_{2,c}$
      scaling (see text for details).
    In both panels,  the  TPC  and  FTPC  points are  
    shifted  horizontally
    relative to one another for clarity purposes.  
    The error bars are statistical.
}
  \label{fig:au200F}
\end{figure}

Figure~\ref{fig:au200F}(a)    compares  the    three-particle
correlations obtained for different charge combinations, as a function
of centrality, when the  third particle is
selected  from the  main  TPC with when it is selected
from  the Forward TPCs.
Assuming  that the second  harmonic of  the  third particle  is
correlated with  the first harmonic  of the first  two particles
via  a  common  correlation  to the  reaction  plane,  the
correlator should  then be proportional to the  elliptic flow of
the third particle.   
On average, the elliptic flow  in the FTPC region
is significantly smaller than that in the TPC
region~\cite{Adams:2004bi}, explaining the different
magnitudes   of  the  three-particle  correlations   shown  in
Fig.~\ref{fig:au200F}(a).

Figure~\ref{fig:au200F}(b)  shows the  three-particle correlator
after  it  has been divided  by  $v_2$  of the  third  particle
according to Eq.~\ref{e3p}. 
Resulting signals are in very good agreement in the two cases.
In  this and subsequent  plots,   
for the elliptic flow of particle $c$ in  the main TPC region
we use estimates obtained from the correlations of particles
in the  main TPC region, $|\eta | <  1.0$, with particles
in the  FTPC, $2.7 < |  \eta | < 3.9$.
These estimates are less  affected by non-flow effects,
compared to elliptic flow  derived from two-particle correlations with
both particles taken from the main TPC.

The shaded  band in Fig.~\ref{fig:au200F}(b) and the subsequent
figures  illustrate the systematic
change in  the results that  occur when different  estimates of
the elliptic flow are  used.  
The upper (in magnitude) limit is obtained with flow from four-particle
correlations and the lower limit from the two-particle cumulant method.
All  elliptic  flow  data have  been  taken  from
  Ref.~\cite{Adams:2004bi,Voloshin:2007af}~\footnote{
In Ref.~\cite{Adams:2004bi,Voloshin:2007af} an estimate
of elliptic flow in the main TPC region, $|\eta | <  1.0$, 
obtained from correlations of particles in this region with those in
FTPCs was denoted as $v_2\{\mathrm{FTPC}\}$; an estimate from
two-particle correlations with both particles in the main TPC as  $v_2\{2\}$.
Elliptic flow from four-particle correlations, denoted
as $v_2\{4\}$, is considered to be least affected by non-flow effects. 
For a review of flow measurements, see~\cite{Voloshin:2008dg}.
}.             
Four-particle cumulant values are  not available  for all
  collision  systems and energies   studied here. 
Therefore in Figs.~\ref{fig:uuv2_200}--\ref{fig:uuv2_AuNpart}, 
we  plot  systematic upper limits obtained with extrapolation of
available data 
assuming that the measurements with FTPC suppress 
only 50\% of the non-flow contribution.  
The magnitude  of  the  elliptic  flow  in  the FTPC  region  was  
estimated  from correlations between  particles 
in the East and  West FTPCs.  
Section~\ref{sec:syst} has further details on the  systematic
uncertainties associated with different $v_2$ estimates.

Results obtained with the  event plane reconstructed with
ZDC-SMD are consistent with those shown in Fig.~\ref{fig:au200F}(b),
though the  statistical errors  on ZDC-SMD results  are about  5 times
larger because the (second order) reaction plane resolution  from ZDC-SMD
is worse. 

Figure~\ref{fig:au200F}(b) shows very good agreement between the same
charge correlations obtained with the third particle in the TPC and
FTPC regions, which supports for such correlations the assumption
 $\corr   \approx
\mean{\cos(\phia +\phib  -2\phi_c)}/v_{2,c} $.
The opposite charge correlations are small in magnitude and 
it is difficult to conclude on validity of the
assumption for such correlations  based only
on results presented in Fig.~\ref{fig:au200F}(b). 
Similarly, in the most peripheral collisions, the statistical
errors are large, which also prohibits making a definite conclusion.

%----------------------------------------------------------
\section{Systematic uncertainties}
\label{sec:syst}

There is one class of uncertainties, related to the question of factorization 
 of Eq.~\ref{e3p}, which would arise if the events contained
a large number of correlated groups of particles such as minijets. 
Even if these ``clusters'' were produced isotropically in azimuth, 
they might contribute to our observable through correlations 
between the particles
used to determine the reaction plane (particle $c$ in Eq. ~\ref{e3p})
and the particles 
($\alpha,\beta$) used to measure the signal.
We consider this  effect in detail in Section~\ref{sec:simulations}. 
As will be shown there, in Cu+Cu and peripheral Au+Au collisions 
this effect could cause opposite charge correlations of the sign and
magnitude we observe, but does not produce the same charge correlations.

We proceed with discussion of the results assuming 
 $\corr   = \mean{\cos(\phia +\phib  -2\phi_c)}/v_{2,c} $
but indicate in all plots the HIJING~\cite{refHIJING}
(default, quenching-off settings) three-particle correlation results. 
The latter can be
considered as an estimate of the systematic uncertainty
from correlations not related to the reaction plane. 
In future high statistic measurements such uncertainty can be
decreased by taking particle $c$ from a rapidity region separated from
particles $\alpha$ and $\beta$.

One dominant systematic uncertainty in the correlator $\corr$ is
due to uncertainty in the elliptic flow measurements of the particle 
used to determine the reaction plane. 
This contributes a fractional uncertainty, on average 
of the order of 15\% and somewhat
larger in most peripheral and most central collisions~\cite{Adams:2004bi}. 

From comparison of the results obtained in different field
configurations, and other studies presented in
Section~\ref{sec:detector_effects} we conclude that 
after acceptance corrections are performed, the remaining 
systematic uncertainties in three-particle correlations due to
detector effects are comparable to or smaller than 
the statistical errors.

We have performed an additional study to estimate the size of possible error
caused by acceptance effects before and after the recentering
correction is applied: we have run simulations in which 
tracks were generated using realistic single particle distributions
but having no correlation except due to elliptic flow. 
An efficiency loss is introduced similar to that of the STAR detector as a
function of azimuth, transverse momentum, and particle charge. 
We then study the effect of distorting the efficiency in additional
and more extreme ways. 
In all of these cases,  after the recentering correction
is applied, the value of $\mean{ \cos(\phia +\phib -2\phi_c)}$ is
zero for all centralities within the statistical 
precision of the study, which is about $3\times 10^{-6}$ for the most
peripheral bin and decreases to less than $10^{-7}$ 
for the most central bin. 
This is many times smaller than 
the measured signal for all centralities in all cases.

Errors in measuring the magnitude of particle momenta make negligible 
contributions to the correlator used in this analysis, which 
uses only measured azimuthal angles. 
It is therefore robust against many 
 systematic errors which are commonly encountered
in the analyses of the STAR data (space charge
distortion errors leading to momentum biases, etc.).

Theoretical treatments of the correlator defined in Eq.~\ref{eq:obs1}
were developed with charged hadrons in mind.
By using cuts (based on specific energy loss)
to suppress the presence of electrons in our sample,
we have verified that this bias is also smaller than 
the statistical errors.

%==================================================
\section{Results}
\label{sec:results}

\begin{figure}[tbp]
 \includegraphics[width=.46\textwidth]{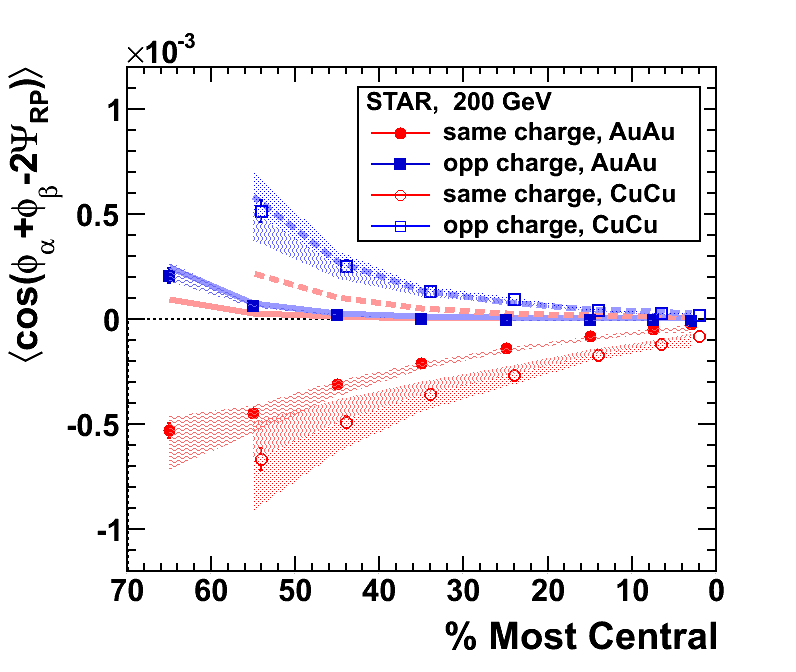}
 \caption{ (Color online)
$\mean{\cos(\phi_a +\phi_\beta -2\psirp) }$ in Au+Au and Cu+Cu
collisions at $\sqrt{s_{NN}}=200$~GeV calculated using Eq.~\ref{e3p}. 
The error-bars show the statistical errors.
The shaded area reflects the uncertainty in the elliptic flow
values used in calculations, 
with lower (in magnitude) limit obtained with elliptic
 flow from two-particle
correlations and upper limit from four-particle cumulants. For
details, see Section~\ref{sec:detector_effects}.
Thick solid (Au+Au) and dashed (Cu+Cu) 
lines represent possible non-reaction-plane dependent contribution from
 many-particle clusters as estimated by HIJING, 
see Section~\ref{sec:backgrounds_sub1}.
}
 \label{fig:uuv2_200}
\end{figure}

Final results presented in this section have been obtained with 
three-particle correlation using Eq.~\ref{e3p} with all three
particles from the pseudorapidity region $|\eta|<1.0$.
Figure~\ref{fig:uuv2_200}
presents the correlator $\corr$ for Au+Au and Cu+Cu
collisions at \snn=200~GeV.
Positive-positive and negative-negative correlations are found
to be the same within statistical errors, see
Fig.~\ref{fig:recenter}(b),
and are combined together as same-charge correlations.
Opposite-charge correlations are relatively smaller 
than same-charge correlations, in agreement with possible
suppression of the back-to-back correlations discussed in the
introduction.
The correlations in Cu+Cu collisions, shown by open symbols,
appear to be larger than the correlations in Au+Au
for the same centrality of the collision.
One reason for this difference may be the difference in number of
participants (or charge multiplicity) in Au+Au and Cu+Cu collisions 
at the same centrality. 
The signal is expected to have a $1/N$ dependence, and at 
the same centrality of the collision the multiplicity is smaller 
in Cu+Cu collisions than in Au+Au.
The difference in magnitude between same and opposite charge correlations
is considerably smaller in Cu+Cu than in Au+Au,
qualitatively in agreement with the scenario of stronger suppression 
of the back-to-back correlations in Au+Au collisions.
In Fig.~\ref{fig:uuv2_200} and below, error bars 
indicate statistical uncertainties.
The shaded bands show the systematic uncertainty associated 
with measurements of elliptic flow 
which are used to rescale the three-particle correlator.
In this section we assume the factorization of correlator Eq.~\ref{e3p}.  
The possible error due to this assumption 
-- which may be large for peripheral bins in the opposite charge
correlation -- is denoted by the thick lines in Fig.~\ref{fig:uuv2_200}
and subsequent figures and is explained in Section~\ref{sec:simulations}.
Other systematic uncertainties have been discussed in section~\ref{sec:syst}.

\begin{figure}[tbp]
 \includegraphics[width=0.5\textwidth]{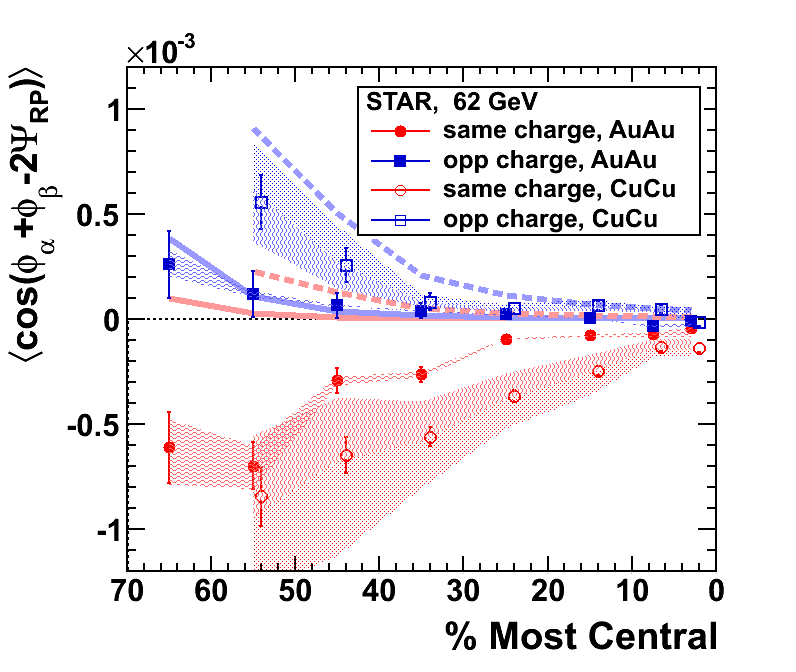}
 \caption{ (Color online)
$\mean{\cos(\phi_a +\phi_\beta -2\psirp) }$ in Au+Au and Cu+Cu
collisions at $\sqrt{s_{NN}}=62$~GeV calculated using Eq.~\ref{e3p}. 
The error-bars indicate the statistical errors.
The shaded area reflects the uncertainty in the elliptic flow
values used in calculations. 
For details, see Section~\ref{sec:detector_effects}.
Thick solid (Au+Au) and dashed (Cu+Cu) 
lines represent possible non-reaction-plane dependent contribution from
 many-particle clusters as estimated by HIJING, 
see Section~\ref{sec:backgrounds_sub1}.
}
 \label{fig:uuv2_62}
\end{figure}

Figure~\ref{fig:uuv2_62} shows results for collisions at \snn=62.4~GeV.
The signal is similar in magnitude, with slightly more pronounced 
opposite-charge correlations compared to those at \snn=200~GeV. 
This is consistent with weaker suppression of opposite-charge correlations 
in the less dense 62~GeV system.

\begin{figure}[tbp]
 \includegraphics[width=0.5\textwidth]{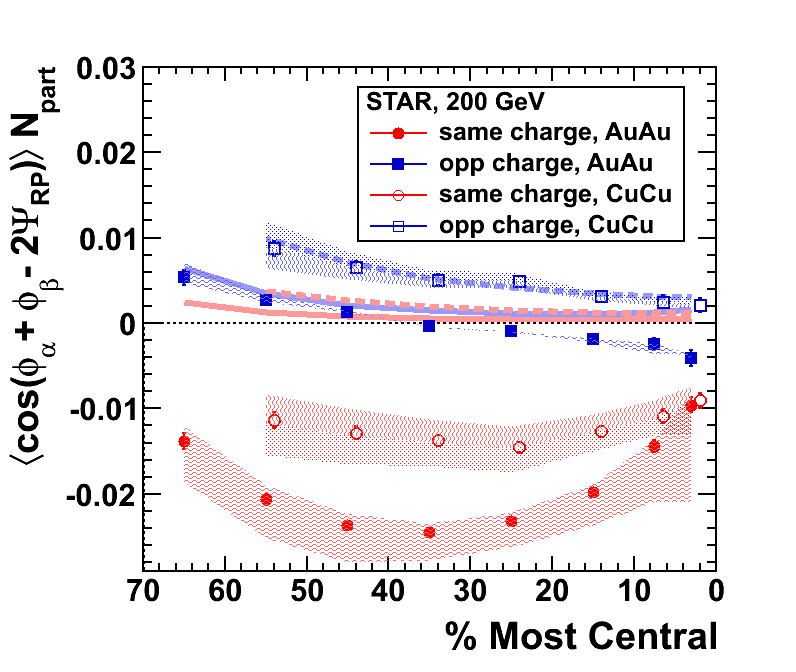}
\centerline{(a)}
 \includegraphics[width=0.5\textwidth]{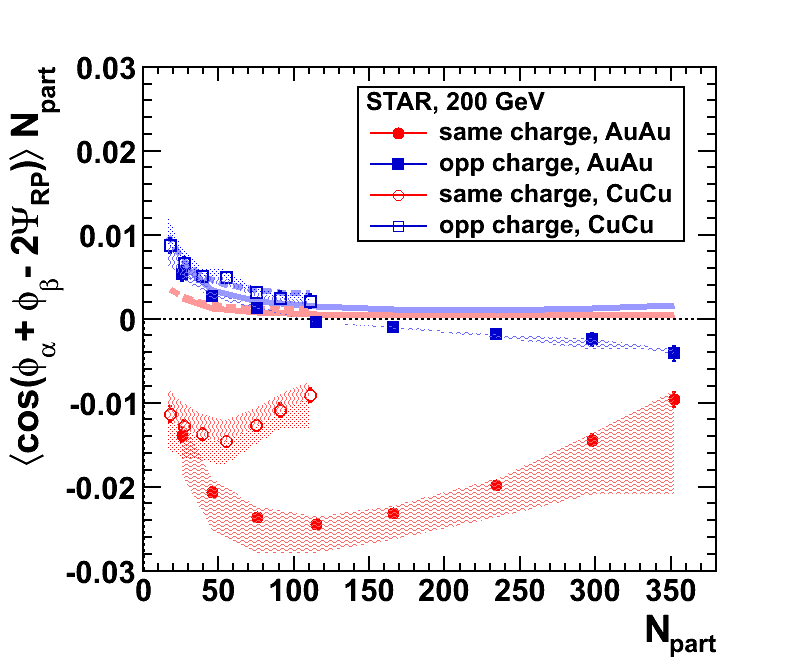}
\centerline{(b)}
 \caption{
(Color online)
 Au+Au and Cu+Cu collisions at \snn=200~GeV. The correlations are  
scaled with the number of participants and are plotted as
function of (a) centrality and (b) number of participants.
The error-bars indicate the statistical errors.
The shaded area reflects the uncertainty in the elliptic flow
values used in calculations. 
For details, see Section~\ref{sec:detector_effects}.
Thick solid (Au+Au) and dashed (Cu+Cu) 
lines represent possible non-reaction-plane
 dependent contribution from
 many-particle clusters as estimated by HIJING, 
see Section~\ref{sec:backgrounds_sub1}.
}
 \label{fig:uuv2_AuNpart}
\end{figure}

The correlations are weaker in more central collisions compared to more
peripheral collisions, which partially can 
be attributed to dilution of correlations
which occurs in the case of particle production from multiple sources. 
To compensate for this effect and to present a more complete picture of
the centrality dependence, we show in Fig.~\ref{fig:uuv2_AuNpart}  
results multiplied by the number of participants. 
The number of nucleon participants is estimated from a Monte-Carlo
Glauber model~\cite{:2008fd}.
Figure~\ref{fig:uuv2_AuNpart}(a) presents the results 
 as a function of centrality, and Fig.~\ref{fig:uuv2_AuNpart}(b) 
 as a function of $N_{part}$.
Smaller correlations in most central collisions are
 expected in the parity violation picture as the magnetic field weakens. 
The same and opposite charge correlations
 clearly exhibit very different behavior.
Figure~\ref{fig:uuv2_AuNpart}(a) demonstrates that 
the same-charge correlations show similar centrality dependencies, as
would be expected if the geometry of the collision is important.
The opposite-charge correlations in Au+Au and Cu+Cu collisions are
 found to be close at similar values of $N_{part}$ in rough
 qualitative agreement with the picture in which their values are mostly
 determined by the suppression of back-to-back correlations.

\begin{figure}[tbp]
 \includegraphics[width=0.5\textwidth]{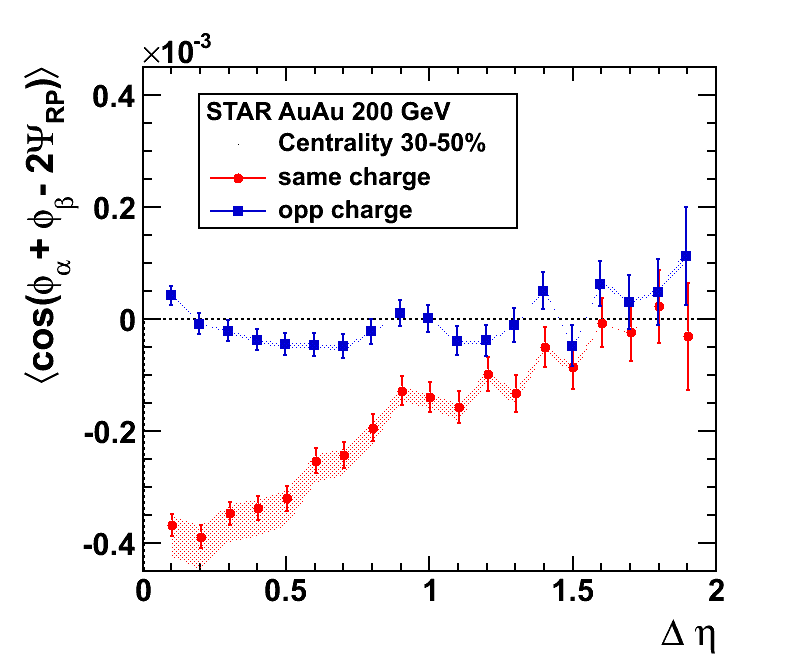} 
	\centerline{(a) }
 \includegraphics[width=0.5\textwidth]{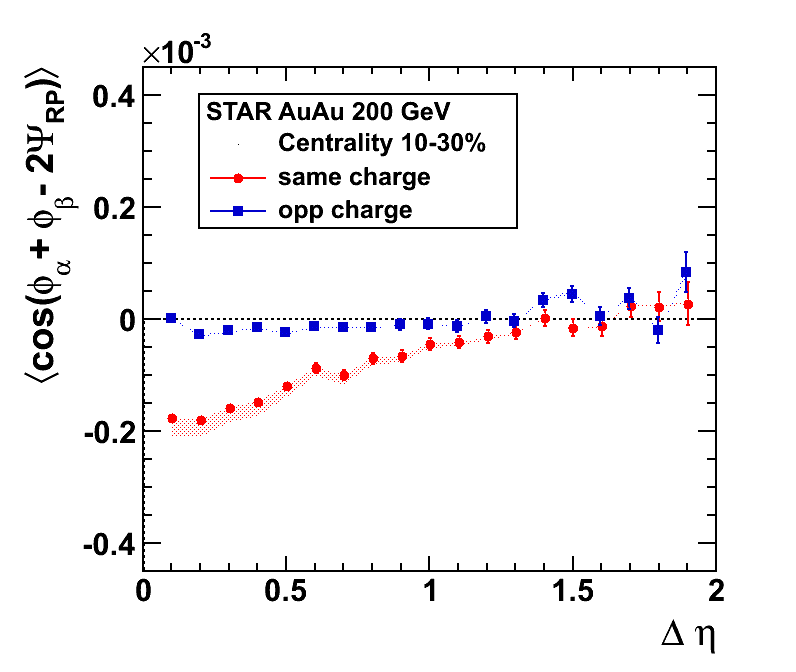} 
	\centerline{(b) }
 \caption{ (Color online)
Au+Au at 200 GeV. The correlations dependence on
pseudorapidity separation
$\Delta\eta=|\eta_\alpha -\eta_\beta|$ for
(a) centrality 30-50\%, and (b) centrality 10-30\%. 
The shaded band indicates uncertainty
  associated with $v_2$ measurements and has been calculated 
using two- and four-particle cumulant results as the limits.
} 
\label{fig:uuv2eta}
\end{figure}

\begin{figure}[tbp]
 \includegraphics[width=0.5\textwidth]{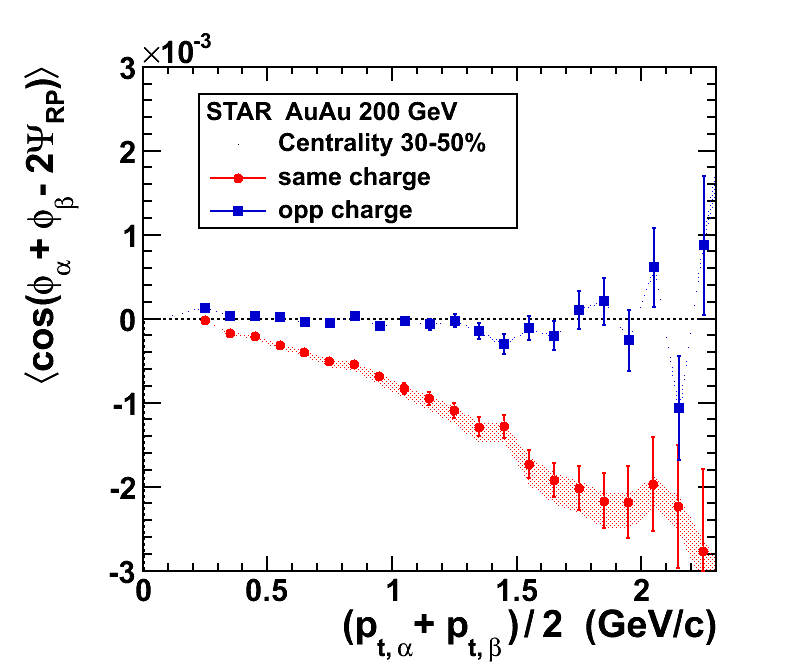} 
	\centerline{(a) }
 \includegraphics[width=0.5\textwidth]{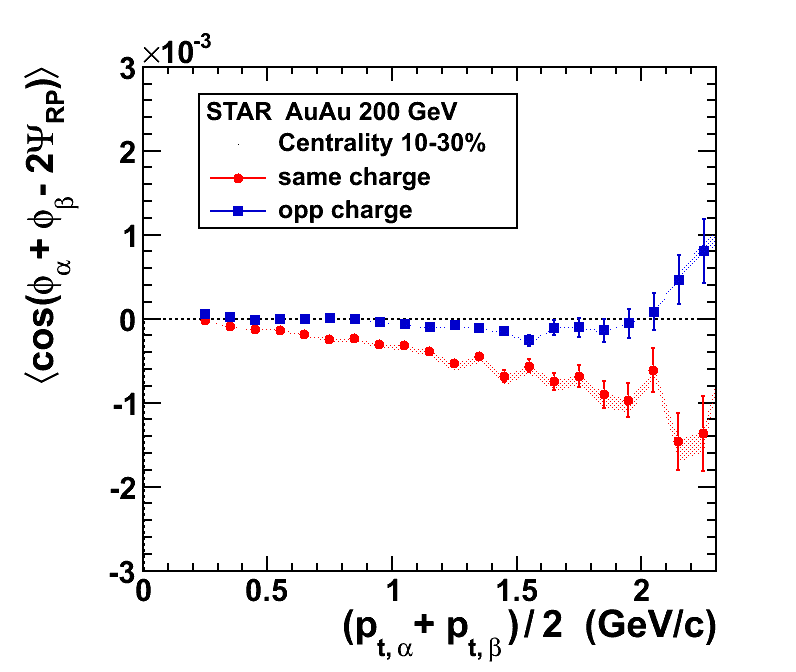}
	\centerline{(b) }
 \caption{ (Color online)
 Au+Au at 200 GeV. The correlations dependence on
$(p_{t,\alpha}+p_{t,\beta})/2$ for (a) centrality 30-50\%, and 
(b) centrality 10-30\%. The shaded band 
has the same meaning as that in Fig.~\ref{fig:uuv2eta}. 
}
 \label{fig:uuv2pt}
\end{figure}

Figure~\ref{fig:uuv2eta} shows the dependence of the signal on the
difference in pseudorapidities of two particles,
 $\Delta \eta = |\eta_\alpha -\eta_\beta|$, for 30-50\% 
and 10-30\% centralities.
The signal has a typical hadronic width of about one unit of
pseudorapidity.
The dependence on $|\eta_\alpha -\eta_\beta|$ has been calculated for
all charged tracks with $0.15<p_t<2.0$~GeV/c.
Figure~\ref{fig:uuv2pt} shows the dependence of the signal on 
the sum of the transverse momentum (magnitudes)  of the two particles for
these same centralities.
Results presented is this figure have no upper $p_t$ cut.
We do not observe the signal concentration in the low $p_t$
region as naively might be expected for \P-violation effects.

\begin{figure}[tbp]
 \includegraphics[width=0.5\textwidth]{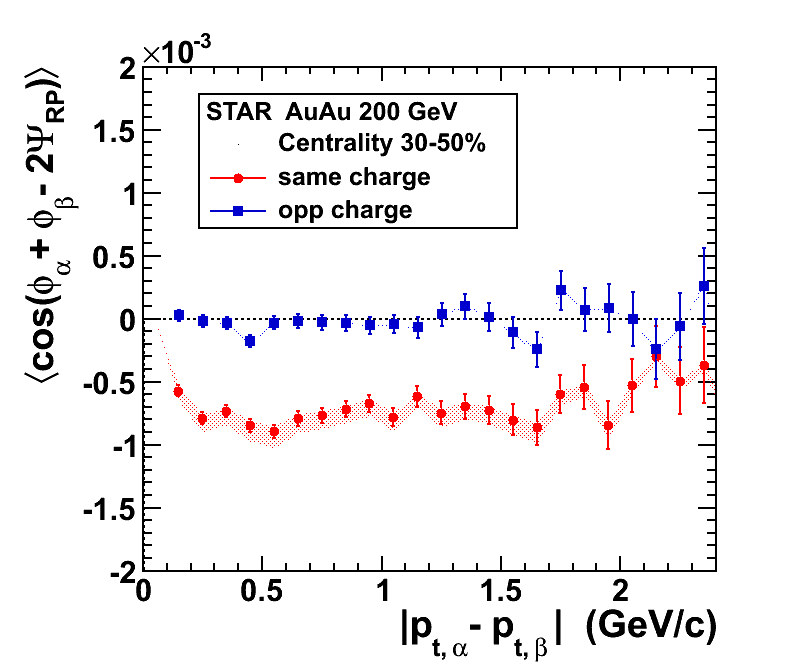} 
	\centerline{(a) }
 \includegraphics[width=0.5\textwidth]{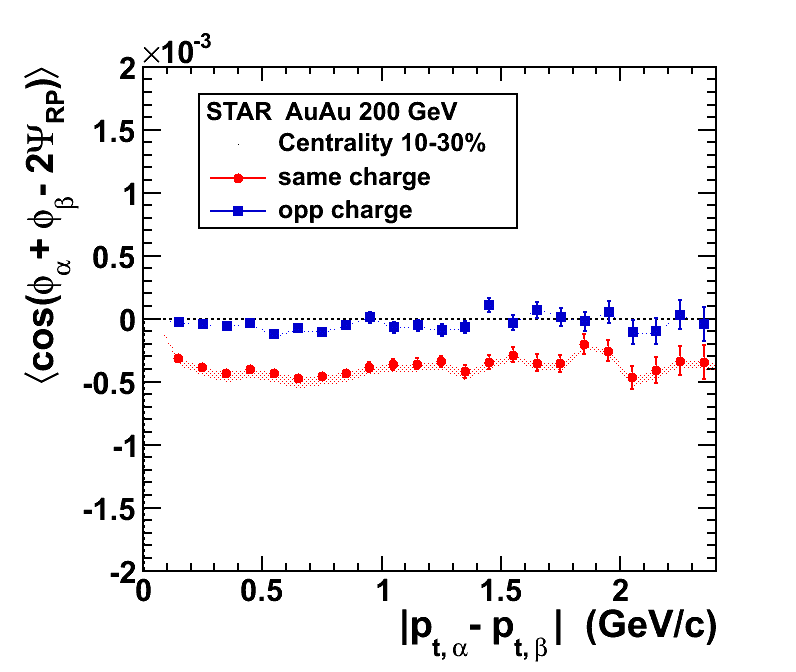}
	\centerline{(b) }
 \caption{ (Color online)	
   Au+Au at 200 GeV. 
	The correlations dependence on
	$|p_{t,\alpha}-p_{t,\beta}|$ for (a) centrality 30-50\%, and 
	(b) centrality 10-30\%. The shaded band  
	has the same meaning as that in Fig.~\ref{fig:uuv2eta}.
}
 \label{fig:uuv2ptDiff}
\end{figure}

Figure~\ref{fig:uuv2ptDiff}  shows the dependence of the signal on the
difference in the magnitudes of the two particle transverse momenta. 
We find that the correlation depends very weakly on
$|p_{t,\alpha}-p_{t,\beta}|$. 
This  excludes quantum interference (HBT) or
Coulomb effects as possible explanations for the signal. 
There are no specific theoretical predictions on this dependence for
the chiral magnetic effect, though naively one expects that the
signal should not extend to large values of $|p_{t,\alpha}-p_{t,\beta}|$.

Finally, the ZDC-SMD detector has good first-order reaction-plane 
resolution. 
For mid-central collisions the resolution, $\mean{\cos(\Psi_1-\psirp)}$,
is of the order of 0.35--0.4. The ZDC-SMD allows us to test the 
first-order (\P-odd) effect of the charge separation 
along the system orbital momentum,
which would correspond to $\mean{a_\alpha} \ne 0$.
In theory this is possible only if the vacuum $\theta \ne 0$.
Our measurements are consistent with zero, averaged over all centralities 
$\mean{a_+}= (-0.1 \pm 1.0) \cdot 10^{-4}$
and $\mean{a_-}= (-1.0 \pm 1.0) \cdot 10^{-4}$.

%------------------------------------------------

\section{Physics backgrounds}
\label{sec:simulations}
\subsection{Reaction-Plane Independent Background}
\label{sec:backgrounds_sub1}

Reaction-plane independent background is caused
by three (or more) 
particle clusters which affect the factorization of Eq.~\ref{e3p}.
With future high statistics data sets, it will be possible to reduce
such backgrounds significantly by determining the reaction plane using
particles far remote in rapidity from the signal particles.

\begin{figure}[ht]
 \includegraphics[width=0.5\textwidth]{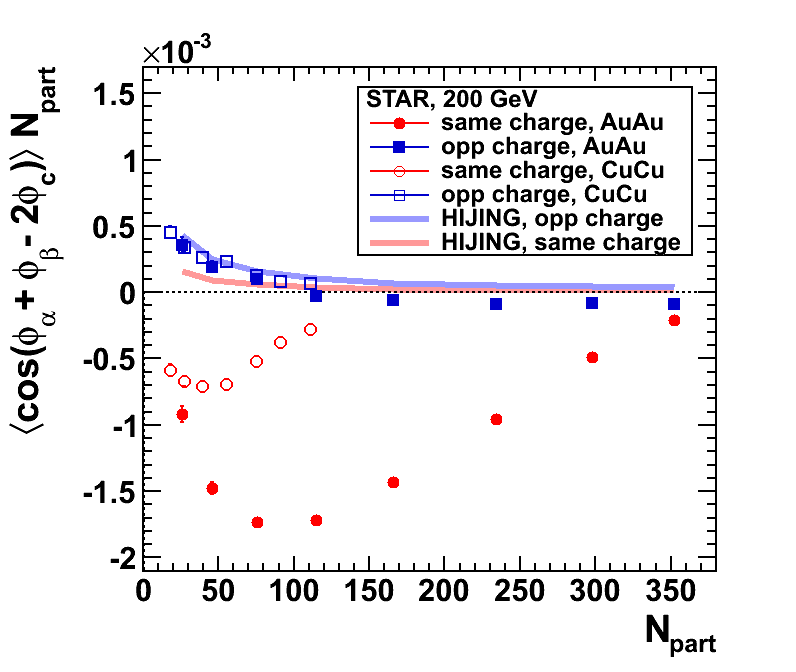}
  \caption{ (Color online)  
    Three-particle correlator in Au+Au and Cu+Cu collisions
  compared to HIJING calculations shown as thick lines. 
  All three particles are taken in the
  main TPC region, $|\eta|<1.0$.
  The correlator has been scaled
  with number of participants and are plotted versus number of
  participants. In this representation HIJING results for Au+Au and
  Cu+Cu collisions coincide in the region of overlap. 
}
  \label{fig:200hj}
\end{figure}

In order to estimate possible contribution to the three-particle
correlator of effects not related to the reaction plane orientation
we use the HIJING~\cite{refHIJING} event
generator which is based on the minijet picture of heavy-ion collisions. 
For all HIJING results presented in this paper we use default,
quenching-off setting.
Figure~\ref{fig:200hj} presents the results for the three-particle
correlator, $\la \cos(\phi_a +\phi_\beta -2\phi_c) \ra$,
 measured in Au+Au and Cu+Cu collisions as a function of
centrality together with HIJING results for the correlations among
three particles from many particle clusters. 
In this figure the most central points correspond to centrality 0--5\%
and the most peripheral to 60--70\% for Au+Au
collisions and 50--60\% for Cu+Cu collisions.
The correlator is scaled with number of
participants for clarity at large centralities, where the signal is
small in magnitude. 
The correlations are shown as a function of the number of
participants because this gives very similar HIJING results 
for Au+Au and Cu+Cu collisions, implying a dependence only on the charged 
particle rapidity density.  
We have separately checked that HIJING results scale  as $N^{-2}$
 as expected for contributions from many particle clusters. 
Figure~\ref{fig:200hj} shows that if this minijet picture in HIJING is
correct, in peripheral collisions the entire
opposite charge signal may be dominated by contributions from
clusters not related to the reaction plane orientation.
The same charge correlations in HIJING are significantly smaller in
magnitude than in data and have opposite sign.  HIJING results for 
three-particle correlations among three particles all of the same charge
are consistent with zero in sharp contrast to the data shown in 
Fig.~\ref{fig:recenter}.

We have also studied such reaction-plane independent backgrounds 
using the event generator
UrQMD~\cite{refRQMD} and found that the predicted contributions 
to both opposite-charge and same-charge correlations are at least a factor
of two lower than those predicted by HIJING.

\subsection{Reaction-Plane Dependent Background}

Unlike those discussed in Sec.~\ref{sec:backgrounds_sub1}, 
reaction-plane dependent physics
backgrounds can not be suppressed by better methods 
of determining the reaction plane.

The correlator $\mean{\cos(\phia+\phib-2\psirp)}$ is a \P-even 
observable and can exhibit a non-zero signal for effects not 
related to \P-violation.
Among those are processes in which particles $\alpha$ and
$\beta$ are products of a cluster (e.g. resonance, jet,
 di-jets) decay, and the cluster itself exhibits elliptic
flow~\cite{Borghini:2001vi,Adams:2003zg} or decays (fragments)
differently when emitted in-plane compared to out-of-plane.

If ``flowing clusters'' are the only contribution to the
 correlator, we can write:
\be & \la \cos(\phi_\alpha + \phi_\beta -2\psirp) \ra =
\nonumber
\\
&
A_{clust} \, \la \cos((\phi_\alpha + \phi_\beta -2\phi_{clust}) 
+2(\phi_{clust}-\psirp)) \ra_{clust} 
\nonumber
\\
& 
=A_{clust} \,
\la \cos(\phi_\alpha + \phi_\beta -2\phi_{clust}) \ra_{clust}
\; v_{2,clust},
\label{eq:resonance}
\ee 
where $\la ... \ra_{clust}$ indicates that the average is performed only
over pairs consisting of two daughters from the same cluster and the
resulting normalization factor is $A_{clust}=N_{\frac{clust}{event}}
N_{\frac{pairs}{clust}}    /    N_{\frac{pairs}{event}}    $.
Equation~\ref{eq:resonance}  assumes that  there is 
no reaction plane
dependence of $\cos(\phi_\alpha + \phi_\beta -2\phi_{clust})$.
The term $ \la \cos(\phi_\alpha + \phi_\beta -2\phi_{clust}) \ra $
is a measure of the azimuthal correlations of decay products
with respect to the cluster azimuth, while $v_{2,clust}$ is
 cluster  elliptic flow.  
In the case of resonance decays,
$\la \cos(\phi_\alpha + \phi_\beta -2\phi_{res}) \ra $ is zero if the
resonance is at rest, and becomes non-zero only due to resonance
motion. 
Estimates of the contribution of ``flowing resonances'',
based on Eq.~\ref{eq:resonance} and reasonable values of resonance
abundances and values of elliptic flow, indicate that they should not
produce a  fake signal.
Given the relative scarcity 
of parents decaying to two same-charge daughters, 
a much smaller magnitude is expected for same charge 
than opposite charge
correlations from this source. 
Kinematic studies demonstrate that it is very difficult for 
the sign  of the correlations observed in the data to be created
in the same-charge correlations without postulating a negative value of
$v_2$ for the resonances or particles from cluster decays. 

\begin{figure}[tbp]
 \includegraphics[width=0.5\textwidth]{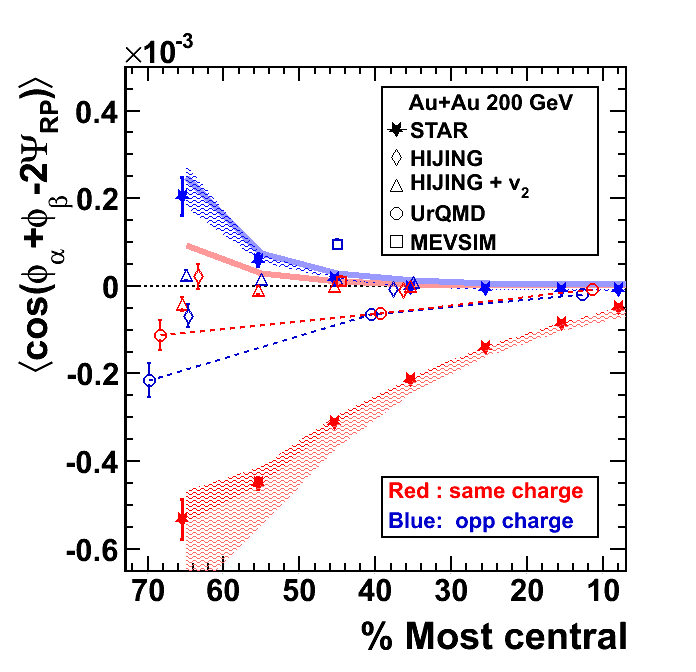}
 \caption{ (Color)
$\corr$ calculated for 200~GeV Au+Au events with
  event generators HIJING (with and without an ``elliptic flow
  afterburner''), UrQMD, and MEVSIM. Blue symbols mark opposite-charge
  correlations, and red are same-charge.  
  Solid stars
  represent the values from the data to facilitate comparison.
  Acceptance cuts of $0.15<p_t<2$~GeV/c and $|\eta|<1.0$ were used
  in all cases. For MEVSIM, HIJING, and UrQMD points the true 
  reaction plane from the generated event was used for $\psirp$.
  Thick solid lighter colored 
lines represent possible non-reaction-plane
 dependent contribution from
 many-particle clusters as estimated by HIJING 
 and discussed in
 Section~\ref{sec:backgrounds_sub1}.
Corresponding estimates from UrQMD are about factor of two smaller.
}
 \label{fig:AuAusimulations}
\end{figure}

To study the contribution from resonances
in greater detail we have carried out simulations using the MEVSIM
event generator~\cite{refMEVSIM}. 
MEVSIM generates particles according to the single particle momentum 
distributions measured at RHIC. 
The only correlations included are an overall bulk
elliptic flow pattern and correlations between daughters of the same
resonance decay (resonances included are $\phi$, $\Delta$, $\rho$,
$\omega$, and $K^{*}$).
MEVSIM  simulation results  are  shown as solid squares  in
Fig.~\ref{fig:AuAusimulations}; the opposite charge correlations are
larger than what is seen in the data, while the same charge
correlations are far smaller in magnitude and of the wrong sign to
match parity violation correlations.  
We conclude that resonances are not responsible for the observed signal.

In addition to contributing to reaction-plane independent background
as discussed in Section ~\ref{sec:backgrounds_sub1}, jets 
are another potential source of reaction-plane dependent
background since their properties
may vary with respect to the reaction plane.
For those jets in a heavy-ion event which include a charged particle of
sufficiently high $p_t$ to act as a trigger particle for a jet analysis, we
may estimate the contribution using the results of previous STAR
jet studies~\cite{Adams:2003kv,Adams:2005ph,Feng:2008an}.  With trigger
transverse momentum values that allow such analysis ($p_t > 3$~GeV/c)
the contribution to $\corr$ is roughly two orders of
magnitude below the same charge signal shown 
in Fig.~\ref{fig:uuv2_200}.  
To extend the study of jet contributions to lower momentum, we rely on event
generators (in particular, HIJING) calculations.

Several correlation measurements from 
RHIC~\cite{Alver:2007wy,Daugherity:2008su}
and earlier measurements at ISR (see review~\cite{Foa:1975eu})
indicate that cluster formation plays an important role in multiparticle
production at high energies.
These clusters, with a size inferred in ~\cite{Alver:2007wy} to
be 2.5--3 charged particles per cluster,
may account for production of a significant fraction of all particles.
Because we have limited information about the nature of these 
clusters we do not
make an estimate of their contribution to the observed correlations.
Our studies indicate that in order
to fake the same-charge correlations observed in the data, 
there should be several types of clusters with some of them having
negative values of elliptic flow.
It is hoped that with a better understanding of the cause and
properties (including charge dependence and $v_2$) of such clusters,
a clearer statement can be made regarding their contributions.

We have also run simulations with several p+p and Au+Au
event generators.  
With PYTHIA~\cite{Sjostrand:2006za} p+p  events we find  that the
correlations in ${\mean{\cos(\phia-\phib)}}$ are significantly smaller
than those seen in Au+Au data when scaled by $1/N$, and are
similar for all charge combinations. 
We add modulation with respect to the reaction plane 
by  adding  $v_2$  through  angular correlations  or  strong
(elliptically modulated) radial flow. 
This way we create non-zero values for $\corr$, albeit with 
correlations different from the data, being always positive
and similar in magnitude for all charge combinations.

Figure~\ref{fig:AuAusimulations} shows 
results for (reaction-plane dependent) physics backgrounds to
$\corr$
calculated with 200~GeV Au+Au events from the event generators
UrQMD~\cite{refRQMD} and HIJING~\cite{refHIJING}.  
Because the modulation of $dN/d\phi$ with 
respect to the reaction plane is smaller in  HIJING than seen in RHIC
data, we also run HIJING  with an added
``afterburner'' which adds elliptic flow using as input $v_2$ values
consistent with STAR measurements at the given centrality.  
Elliptic flow is introduced by the ``shifting''
method~\cite{Poskanzer:1998yz}, which preserves other correlations
that exist in the model. 
Figure~\ref{fig:AuAusimulations} shows
that no generator gives qualitative agreement with the data; the model 
values of $\corr$ are significantly smaller in magnitude than what is
seen in the data, and the correlations calculated in these
models tend  to be very  similar for same and  opposite charge
correlations.
 
These models do not match the correlations for
${\mean{\cos(\phia-\phib)}}$ that are seen in the data either,
as shown in Fig.~\ref{fig:uvAu200}. 
HIJING predicts very similar same and
opposite charge correlations that are much smaller in magnitude than
seen in the data. 
UrQMD overestimates the same charge correlations. 
It predicts opposite charge correlations that are much
smaller in magnitude and opposite in sign from the data. 
This points to the
need for better modeling of two-particle correlations to give
quantitatively meaningful comparisons for $\corr$.

In Fig.~\ref{fig:AuAusimulations} we connect UrQMD points by dashed
lines to illustrate that the ``reference line'' for strong parity
correlations might be not at zero. 
In this particular case of UrQMD, both same
and opposite-charge correlations have values below zero.
Note that the same-charge correlations sit somewhat above 
the opposite-charge correlations,
opposite to the expectation from local parity violation.

Directed flow, which on average is zero in a symmetric pseudorapidity
interval, can contribute to the correlator $\corr$ via 
flow fluctuations. 
This  effect  is of  the  opposite sign,  see Eq.~\ref{eq:v-a}, 
and is similar for different charge combinations unlike the signal. 
If one assumes
that the amplitude of the fluctuations is of the same order of magnitude
as the maximum directed flow in the pseudorapidity interval under
study, then the flow fluctuation contribution is no more than 10$^{-5}$
for centrality 
30-60\%, significantly smaller than the observed signal.

Global polarization
of hyperons~\cite{Liang:2004ph,Voloshin:2004ha}, the phenomenon of the
polarization of secondary produced particles along the direction of
the system's angular momentum, may also contribute to 
the correlator Eq.~\ref{eq:obs1}
via \P-odd weak decays. This effect could lead to a charge asymmetry
with respect to the reaction plane, which is always pointing in the
same direction relative to the orientation of the angular momentum.
Our main analysis based on the reaction plane reconstructed from the
elliptic flow does not distinguish the direction of the angular
momentum, and is susceptible to this effect. But as we pointed out in
section~\ref{sec:results}, our measurement of charge separation along
the system orbital angular momentum is zero based on the first-order  
reaction  plane reconstructed  in  the ZDC-SMD.  Global
polarization has also been found to be consistent with
zero, $P_{\Lambda, \bar{\Lambda}}<0.02$~\cite{Abelev:2007zk}.

%===========================================================
\section{Summary}

An analysis using three-particle correlations that are directly
sensitive to the \P-violation effects in heavy-ion collisions has
been presented for Au+Au and Cu+Cu collisions at $\sqrt{s_{NN}}$=200
and 62~GeV.
The results are reported for different particle charge combinations 
as a function of collision centrality, particle
separation in pseudorapidity, and particle transverse momentum.
Qualitatively the results agree 
with the magnitude and gross features
of the theoretical predictions for local \P-violation in heavy-ion
collisions, except that
 the signal persists to higher transverse
momenta than expected~\cite{Kharzeev:2007jp}.
The particular observable used in our analysis is
\P-even and might be sensitive to non-parity-violating effects. 
So far, with the systematics checks discussed in this paper, 
we have not identified  effects that would explain the
observed same-charge correlations.
The observed signal cannot be described by the background models that
we have studied 
(HIJING, HIJING+$v_2$, UrQMD, MEVSIM), 
which span a broad range of hadronic physics.

A number of future experiments and analyses are naturally suggested by
these results. 
One of them is the study of the correlation dependence 
 on the energy of the colliding ions. 
The charge separation effect is expected to depend strongly 
on the formation of a quark-gluon plasma~\cite{Kharzeev:2007jp}, 
and the signal might be greatly suppressed or completely absent
at an energy below that at which a quark-gluon plasma can be formed.

Improved theoretical calculations of the expected signal and potential 
physics backgrounds in high energy heavy ion collisions are essential 
to understand whether or not the observed signal is due to local 
strong parity violation, and to further experimental study of this phenomenon.

\section*{Acknowledgments}  

We thank D.~Kharzeev for discussions of the local
strong parity violation phenomenon and its experimental signatures. 
We thank the RHIC Operations Group and RCF at BNL, and the NERSC Center 
at LBNL and the resources provided by the Open Science Grid consortium 
for their support. This work was supported in part by the Offices of NP 
and HEP within the U.S. DOE Office of Science, the U.S. NSF, the Sloan 
Foundation, the DFG cluster of 
excellence `Origin and Structure of the Universe', 
CNRS/IN2P3, RA, RPL, and EMN of France, 
STFC and EPSRC of the United Kingdom, FAPESP 
of Brazil, the Russian Ministry of Sci. and Tech., the NNSFC, CAS, MoST, 
and MoE of China, IRP and GA of the Czech Republic, FOM of the 
Netherlands, DAE, DST, and CSIR of the Government of India,
the Polish State Committee for Scientific Research,  and the Korea Sci. 
\& Eng. Foundation.

%-------------------------------------------------------

\end{document}